\documentclass[journal]{IEEEtran}
\ifCLASSINFOpdf
\else
\fi

\usepackage{cite}
\usepackage{amsmath,amssymb,amsfonts} \allowdisplaybreaks[4]
\usepackage{textcomp}
\usepackage{subfigure}
\usepackage{stfloats}
\usepackage{multirow}
\usepackage{algorithm}
\usepackage{algorithmic}
\usepackage{graphicx}
\usepackage{setspace}
\usepackage{color}
\usepackage{bm}
\usepackage{mathrsfs}
\usepackage{ntheorem} 
 
\newtheorem{assumption}{Assumption} 
\newtheorem{theorem}{Theorem} 
 
\newtheorem{Proposition}{Proposition}
\newtheorem{lemma}{Lemma} 

\newenvironment{proof}{{\indent \indent \it Proof:}}{\hfill $\square$\par}

\begin{document}
%
\title{Transforming Time-Varying to Static Channels: \\
The Power of Fluid Antenna Mobility}
%
%
%

\author{Weidong~Li,~Haifan~Yin,~\IEEEmembership{Senior~Member,~IEEE},~Fanpo~Fu,~Yandi~Cao~and~M\'{e}rouane~Debbah,~\IEEEmembership{Fellow,~IEEE}
\thanks{W. Li, H. Yin, F. Fu and Y. Cao are with the School of Electronic Information and Communications, Huazhong University of Science and Technology, 430074 Wuhan, China (e-mail: weidongli@hust.edu.cn, yin@hust.edu.cn, fanpofu@hust.edu.cn, ydcao@hust.edu.cn).}
\thanks{M. Debbah is with KU 6G Research Center, Khalifa University of Science and Technology, P O Box 127788, Abu Dhabi, UAE (email: merouane.debbah@ku.ac.ae) and also with CentraleSupelec, University Paris-Saclay, 91192 Gif-sur-Yvette, France.}
\thanks{The corresponding author is Haifan Yin.}
\thanks{This work is supported by the National Natural Science Foundation of China under Grant 62071191.}}


\maketitle

\begin{abstract}
This paper addresses the mobility problem with the assistance of fluid antenna (FA) on the user equipment (UE) side.
We propose a matrix pencil-based moving port (MPMP) prediction method, which may transform the time-varying channel to a static channel by timely sliding the liquid.
Different from the existing channel prediction method, we design a moving port selection method, which is the first attempt to transform the channel prediction to the port prediction by exploiting the movability of FA.
Theoretical analysis shows that for the line-of-sight (LoS) channel, the prediction error of our proposed MPMP method may converge to zero, as the number of BS antennas and the port density of the FA are large enough. 
For a multi-path channel, we also derive the upper and lower bounds of the prediction error when the number of paths is large enough. 
When the UEs move at a speed of 60 or 120 km/h, simulation results show that, with the assistance of FA, our proposed MPMP method performs better than the existing channel prediction method.

\end{abstract}


\IEEEpeerreviewmaketitle

\section{Introduction}
\IEEEPARstart{T}{he} fluid antenna (FA) (or ``movable antenna" (MA)) is a new promising technique to enhance the future communication system \cite{Khammassi23TWC,21TWC-KKW,23TWC-KKW,24TWC-KKW}.
Different from the fixed antenna, FA introduces a software-controllable liquid-metal structure that freely switches the liquid to any predetermined locations (referred to as ``ports”) within a given space.
By optimizing the port selection, the FA system has the potential to provide additional spatial degrees of freedom (DoF), mitigate interference, and enhance the channel capacity \cite{Khammassi23TWC}.

Some literature has evaluated the superior performance of FA over the fixed antenna. The authors in \cite{21TWC-KKW} derive the upper bound of the outage probability (OP) of the FA system and prove that as the number of ports is large enough, the system with an FA outperforms the maximum ratio combining (MRC) system with multiple fixed antennas.
Apart from OP, the work in \cite{23TWC-KKW} derives the approximated closed-form expression of diversity gain and determines the minimal number of ports that ensures the single FA system outperforms the single fixed antenna system.
Furthermore, in \cite{24TWC-KKW}, the authors derive the closed-form lower bound of the OP and prove that for an FA with a fixed length, the OP will decrease if the number of ports is large enough.
The above works show that by reshaping the length and the number of ports, the FA performs better than the fixed antenna.

On this basis, some researchers try to achieve better performance by port selection of the FA system.
The paper \cite{24WCL-Qin} investigates FA port optimization to design the BS beamforming matrix, which may achieve the minimum signal-to-interference-plus-noise ratio (SINR). 
The work of \cite{24Arxiv-ZhangRui} considers a set of 6D MA system, and maximizes the network capacity by position optimization.
To maximize the received signal power, the work in \cite{24WCL-ZhangRui} introduces the graph theory and proposes a sequential update algorithm to select the location of MA.
However, these works assume the channel state information (CSI) for all ports of the FA is known. 
In fact, the perfect CSI is difficult to acquire, and the base station (BS) usually estimates CSI from the uplink (UL) channel.

Recently, some works are studying the FA channel estimation.
The authors in \cite{24TWC-ZhangRui} design an orthogonal matching pursuit (OMP)-based channel estimation scheme, which estimates the angles and complex coefficients, and reconstructs the FA channel.
The paper \cite{23TCom} proposes a sequential linear minimum mean-squared error (LMMSE)-based channel estimation method.
In \cite{24CL-KKW}, the authors propose two FA channel estimation schemes, i.e., the least squares (LS) method and the compressed sensing (CS)-based method. Unlike the LS method, which requires the prior FA CSI of all ports, the CS-based method only needs the CSI of several ports to estimate the parameters, e.g., the number of paths, angles, and path gains. Based on the estimated parameters, they reconstruct the FA channel.
The paper \cite{24WCNC-Dai} models the FA channel of all ports as a stochastic process and proposes a successive Bayesian reconstructor method to estimate the channel.
Machine learning is also applied to estimate the FA channel \cite{24Arxiv-Ji}.
However, few literature studies the mobility problem, and the proposed channel estimation schemes may not achieve the expected performance in the practical mobility scenarios.

The mobility problem (or named “the curse of mobility”) brings significant performance loss in the wireless communication system. The main reasons of the mobility problem are CSI delay and the user equipment (UE) movement, where CSI delay is the time interval between the BS estimating the CSI and the downlink (DL) precoding. The UE movement brings non-negligible Doppler effect, which makes the channel time-varying and leads to the outdated CSI. Channel prediction is an effective method to address the mobility problem, which estimates CSI from the UL channel on the BS side and predicts the future DL channel based on historical samples. 

To address the mobility problem, the work in \cite{Yin20JSAC} proposes a Prony-based angular delay domain (PAD) channel prediction method, and proves that as the number of BS antennas is large enough, the PAD method achieves an error-free prediction. 
In \cite{Li24TWC}, the authors propose a wavefront transformation-based matrix pencil (WTMP) method to predict the ELAA channel. It designs a wavefront transformation matrix that transforms the near-field channel and makes it closer to the far-field channel. 
The above methods have addressed the mobility problem in a fixed antenna system, and there is still potential to exploit the FA system to improve the communication performance further. 
The authors in \cite{24WCL-TimeVarying} iteratively optimize the FA port to maximize system performance by the multi-armed bandit learning framework. However, the learning framework needs much training time, and the generalization may need to be enhanced.

To fill the above gaps, we propose a matrix pencil-based moving port (MPMP) prediction method with the assistance of FA.
Unlike the existing prediction methods in \cite{Yin20JSAC} and \cite{Li24TWC}, our proposed MPMP method transforms the channel prediction to the port prediction.
In the traditional fixed antenna communication system, e.g., multiple-input multiple-output (MIMO) and massive MIMO, the channel is time-varying in the mobility scenarios, and the huge computational overhead brings a large CSI delay and increases the burden of channel prediction. 
With the assistance of FA, we may transform the time-varying channel to make it close to the static channel by sliding the liquid.
The moving port prediction question includes two sub-questions, i.e., the parameter estimation question on the BS side and the time-varying port prediction question on the UE side. In other words, the BS estimates the channel parameters from the UL pilot, and then transmits them to the UE. Finally, the UE predicts the future FA ports.
Different from the existing methods where the BS estimates the CSI during different coherence intervals, our MPMP method requires that the BS estimates the channel parameters once in a stationary time - the time duration over which the multipath angles and delays, which is much larger than the channel coherence time \cite{FraunhoferIIS:RP-193072}.

In this paper, leveraging the matrix pencil (MP) algorithm, we first estimate the channel parameters, e.g., the number of paths, path gains, Doppler, and angles.
Then, we reconstruct the channel with the estimated parameters and, on this basis, construct the optimization problem.
Next, we design a port prediction method.
Finally, we reconstruct the channel and keep it static by sliding the liquid to the selected port.
To the best of our knowledge, our proposed MPMP prediction method is the first attempt to predict the FA port in the mobility scenarios.

The contributions of this paper are summarized as follows:

\begin{itemize}
\item
We propose an MPMP prediction method to address the mobility problem with the assistance of FA, which transforms the time-varying channel into a static channel by sliding the liquid. Unlike the existing prediction method, the MPMP method transforms the channel prediction to the port prediction, reducing the computation overhead on the BS side.
Simulation results show that our MPMP outperforms the traditional methods.

\item
We derive the lower bound of the FA length such that the FA contains at least one global optimal port.
We also derive the maximum sliding speed, which is related to the angles, Doppler, wavelength, and time sampling interval.

\item
We prove that for a line-of-sight (LoS) channel, if the FA length is large enough, the prediction mean square error (MSE) of the MPMP method converges to zero, providing that the number of BS antennas and the port density are large enough.

\item
Furthermore, we analyze the prediction MSE of the MPMP method under a multi-path FA channel. We derive the MSE lower and upper bounds, when the FA length, the number of BS antennas and the number of paths are large enough.

\end{itemize}

This paper is organized as follows: We first introduce the channel model in Sec. \ref{sec:system model}. Then, Sec. \ref{sec:MPMP prediction method} describes our proposed MPMP prediction method. After that, in Sec. \ref{sec:Performance analysis}, the performance of the MPMP method is analyzed. Finally, we give the simulation results in Sec. \ref{sec:Simulations} and conclude the paper in Sec. \ref{sec:Conclusion}.

Notations: We use boldface to represent vectors and matrices.
${({\bf{X}})^T}$ and ${({\bf{X}})^H}$ denote the transpose and conjugate transpose of a matrix ${\bf{X}}$. 
${\left\| \cdot \right\|_2}$ stands for the $L_2$ norm of a vector or a complex value, and $\angle (\cdot)$ denotes the angle of a complex value.
$r \{  \cdot \}$ represents the rank of a matrix.
$E\{  \cdot \} $ is the expectation operation. 
${\bf{X}} \otimes {\bf{Y}}$ is the kronecker product of ${\bf{X}}$ and ${\bf{Y}}$.
${J_0}(x)$ is the Bessel function and ${I_0}(x)$ is the Bessel function of an imaginary argument.
$\left[ x \right]$ denotes the rounding operation of the number $x$.
$lcm(x,y)$ is the least common multiple of $x$ and $y$.

\section{Channel Model}\label{sec:system model}
We consider a time division duplexed (TDD) system where the BS has multiple antennas to serve multiple UEs.

\begin{figure}[!htb]
\centering
\includegraphics[width=2.9in]{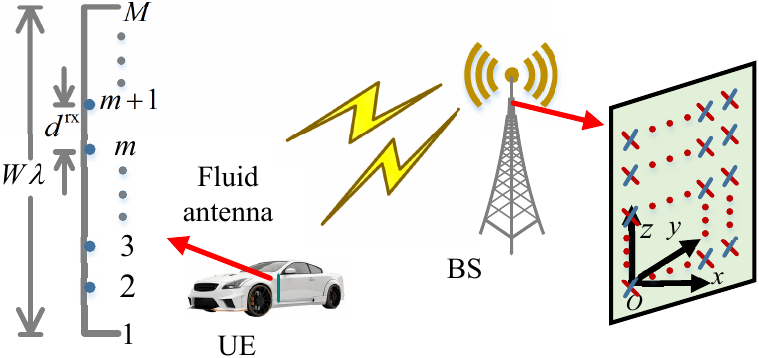}
\caption{The DL wireless communication system}
\label{fig:SystemModel}
\end{figure}

Fig. \ref{fig:SystemModel} gives the DL wireless communication system, where the BS has a ${N_h} \times {N_v}$ uniform planar array (UPA) and the number of BS antennas is ${N_t} = {N_h}{N_v}$.
The UE is equipped with an FA, and the FA contains $M$ ports.
The FA length is ${W\lambda }$, where $\lambda  = \frac{{\rm{c}}}{{{f_c}}}$ is the wavelength with ${{\rm{c}}}$ and ${{{f_c}}}$ being the speed of light and the carrier frequency, respectively.
The spacing ${d_{v}^{\rm{rx}}}$ between two neighboring ports is calculated by ${d_{v}^{\rm{rx}}} = \frac{{\lambda }}{{\rho}}$, where $\rho  = \frac{{M - 1}}{W}$ is defined as the density of ports.
For each UE, the port is selected independently.
Without loss of generality, we assume that the UE selects one port at one slot, by sliding liquid. 
The BS antenna array is located on the $yOz$ plane.
We define the lower location of the leftmost antenna as the coordinate origin, i.e., $[0,0,0]^T$.
The location vector of the $n$-th BS antenna is: 
\begin{equation}
\begin{array}{l}
{\bf{d}}_{n}^{\rm{tx}} = {[0,{d_{h}}(n_h - 1),{d_{v}}(n_v - 1)]^T},
\end{array}
\!\label{LocationFluidAntenna}
\end{equation}
where $d_h$ and $d_v$ are the horizontal and vertical spacings.
$n_h$ and $n_v$ are the horizontal and vertical indices of the $n$-th antenna.
The UE FA ports are arranged along the $z$ axis.
We define the lowest port as the coordinate origin of the UE antenna.
The location vector of the $m$-th port is: 
\begin{equation}
\begin{array}{l}
{\bf{d}}_{m}^{\rm{rx}} = {[0,0,{d_{v}^{\rm{rx}}}(m - 1)]^T},
\end{array}
\!\label{LocationFluidAntennaBS}
\end{equation}
where  $1 \le {m} \le M$.
The spherical unit vectors of the UE and BS are defined as:
\begin{equation}
{\bf{ r}}^{\rm{rx}} = \left[ \begin{array}{l}
\sin {\theta _{{\rm{EOA}}}}\cos {\phi _{{\rm{AOA}}}}\\
\sin {\theta _{{\rm{EOA}}}}\sin {\phi _{{\rm{AOA}}}}\\
\ \ \ \ \ \ \cos {\theta _{{\rm{EOA}}}}
\end{array} \right],
\!\label{SphericalVectorUE}
\end{equation}
\begin{equation}
{\bf{r}}^{\rm{tx}} = \left[ \begin{array}{l}
\sin {\theta _{{\rm{EOD}}}}\cos {\phi _{{\rm{AOD}}}}\\
\sin {\theta _{{\rm{EOD}}}}\sin {\phi _{{\rm{AOD}}}}\\
\ \ \ \ \ \ \cos {\theta _{{\rm{EOD}}}}
\end{array} \right],
\!\label{SphericalVectorBS}
\end{equation}
where ${\theta _{{\rm{EOA}}}}$, ${\phi _{{\rm{AOA}}}}$, ${\theta _{{\rm{EOD}}}}$ and ${\phi _{{\rm{AOD}}}}$ denote the elevation angle of arrival (EOA), the azimuth angle of arrival (AOA), the elevation angle of departure (EOD), and the azimuth angle of departure (AOD).
The ranges of the angles are ${\phi _{{\rm{AoD}}}}, {\phi _{{\rm{AoA}}}} \in (-\pi,\pi ]$ and ${\theta _{{\rm{EoD}}}}, {\theta _{{\rm{EoA}}}} \in [0,\pi ]$.
The Doppler is defined by ${\omega } = \frac{{{{({{\bf{r}}^{\rm{rx}}})}^T}{\bf{v}}}}{\lambda }$, where ${\bf{v}}$ is the UE velocity vector.

We add the superscripts of ``LoS" and ``non-line-of-sight (NLoS)" to denote the spherical unit vectors, angles and the Doppler of the LoS and NLoS paths, respectively.

Let ${h_{u,n,m}}(t)$ denote the channel between the $n$-th BS antenna and the $m$-th port of FA at the $u$-th UE side, which is modeled as \cite{3GPPR16}
\begin{equation}
\begin{array}{l}
{h_{u,n,m}}(t) = \alpha ^{\rm{LoS}} h_{u,n,m}^{{\rm{LoS}}}(t) + \alpha ^{\rm{NLoS}} h_{u,n,m}^{{\rm{NLoS}}}(t)
\end{array},
\!\label{3GPPModel}
\end{equation}
where $\alpha ^{\rm{LoS}} = \sqrt {\frac{{{K_R}}}{{1 + {K_R}}}}$ and $\alpha ^{\rm{NLoS}} = \sqrt {\frac{{{1}}}{{1 + {K_R}}}}$ with ${K_R}$ being the Ricean factor.
$ h_{u,n,m}^{{\rm{LoS}}}(t)$ and $h_{u,n,m}^{{\rm{NLoS}}}(t)$ denote the channels of the LoS path and the NLoS path, respectively:
\begin{equation}
\begin{array}{l}
h_{u,n,m}^{{\rm{LoS}}}(t) = {e^{\frac{{j2\pi {{({{\bf{r}}^{\rm{rx},{\rm{LoS}}}})}^T}{\bf{d}}_{m}^{\rm{rx}}}}{\lambda }}} {e^{j2\pi {\omega ^{{\rm{LoS}}}}t}}\\
\ \ \ \ \ \ \ \ \ \ \ \ \ {e^{j2\pi f{\tau ^{\rm{LoS}}}}}{e^{\frac{{j2\pi {{({{\bf{r}}^{\rm{tx},{\rm{LoS}}}})}^T}{{\bf{d}}_n^{\rm{tx}}}}}{\lambda }}},
\end{array}
\!\label{Channel-LoS}
\end{equation}
and 
\begin{equation}
\begin{array}{l}
h_{u,n,m}^{{\rm{NLoS}}}(t) = \sum\limits_{p = 1}^P {} \beta _p^{{\rm{NLoS}}}{e^{\frac{{j2\pi {{({\bf{r}}_p^{\rm{rx},{\rm{NLoS}}})}^T}{\bf{d}}_{m}^{\rm{rx}}}}{\lambda }}}{e^{j2\pi \omega _p^{{\rm{NLoS}}}t}}\\
\ \ \ \ \ \ \ \ \ \ \ \ \ \ \ \ {e^{j2\pi f{\tau _p^{\rm{NLoS}}}}}{e^{\frac{{j2\pi {{({\bf{r}}_p^{\rm{tx},{\rm{NLoS}}})}^T}{{\bf{d}}_{n}^{\rm{tx}}}}}{\lambda }}},\end{array}
\!\label{Channel-NLoS}
\end{equation}
where ${\tau ^{\rm{LoS}}}$ and ${\tau _p^{\rm{NLoS}}}$ are the delays of the LoS path and the $p$-th NLoS path. $f$ is the frequency and $P$ is the number of the NLoS paths.
We assume that there are $S$ clusters in the propagation space.
The $s$-th cluster contains $P_s$ propagation paths.
We obtain $P =\sum\limits_{s = 1}^S P_s$.
Denote the $s$-th cluster power by ${K_s}'$.
According to \cite{3GPPR16}, the amplitude of the $p$-th NLoS path $\beta _p^{{\rm{NLoS}}}$ belonging to the $s$-th cluster is calculated by:
\begin{equation}
\begin{array}{l}
{\beta _p^{{\rm{NLoS}}}} = \sqrt {\frac{{{K_s}'}}{{{P_s}\sum\limits_{s = 1}^S {{K_s}'} }}}
\end{array}.
\!\label{PathGains}
\end{equation}

For notational simplicity, we will drop the subscript of ``$u$" in the following, and $\alpha ^{\rm{LoS}}$ and $\alpha ^{\rm{NLoS}}$ are abbreviated as $\alpha _p$. ${\beta _p^{{\rm{NLoS}}}}$ is abbreviated as ${\beta _p}$.

The 3-D steering vector ${\bf{a}}({\theta _p},{\phi _p}) \in {{\mathbb{C}}^{N_t \times 1}}$ is:
\begin{equation}
\begin{array}{l}
{\bf{ a}}({\theta _p},{\phi _p}) = {{\bf{a}}_h}({\theta _p},{\phi _p}) \otimes {{\bf{a}}_v}({\theta _p})   ,
\end{array}
\!\label{3-DSteeringVector}
\end{equation}
where
\begin{equation}
\begin{array}{l}
{{\bf{a}}_h}({\theta _p},{\phi _p}) = {\left[ {1, \cdots ,{e^{j\frac{{2\pi }}{\lambda }\sin {\theta _p}\sin {\phi _p}{d_h}({N_h} - 1)}}} \right]^T},
\end{array}
\!\label{SteeringVector-h}
\end{equation}
\begin{equation}
\begin{array}{l}
{{\bf{a}}_v}({\theta _p}) = {\left[ {1, \cdots ,{e^{j\frac{{2\pi }}{\lambda }\cos {\theta _p}{d_v}({N_v} - 1)}}} \right]^T}.
\end{array}
\!\label{SteeringVector-v}
\end{equation}
We denote the channel between all BS antennas and the UE FA at time $t$ by
\begin{equation}
\begin{array}{l}
{{{\bf{ h}}}_{m}}(t) = {\left[ {{h_{1,{m}}}(t), \cdots ,{h_{{N_t},{m}}}(t)} \right]^T} = {{{\bf{A}}}}{{\bf{ C}}_{m}}(t),
\end{array}
\!\label{ChannelAllAntenna}
\end{equation}
where ${{\bf{ A}}} \in {{\mathbb{C}}^{N_t \times (P+1)}}$ contains the 3-D steering vectors of all paths:
\begin{equation}
\begin{array}{l}
{{\bf{ A}}} = \left[ {{\bf{ a}}({\theta _1},{\phi _1}), \cdots ,{\bf{ a}}({\theta _{P+1}},{\phi _{P+1}})} \right].
\end{array}
\!\label{AuULAChannel}
\end{equation}
The matrix ${{\bf{C}}_{m}}(t) \in {{\mathbb{C}}^{(P+1) \times 1}}$ is:
\begin{equation}
\begin{array}{l}
\!\!\!{{\bf{C}}_{m}}(t) = {\left[ { {c_{1,m}}{e^{j2\pi \omega _1 t}}, \cdots ,{c_{P,m}}{e^{j2\pi \omega _{P+1} t}}} \right]^T},
\end{array}
\!\!\!\!\label{CuFluidChannel}
\end{equation}
where ${c_{p,m}} = {c_{p}} {e^{j\frac{{2\pi }}{\lambda }\cos {\theta _p^{\rm{rx}}}d_v^{{\rm{rx}}}(m - 1)}}$ with ${c_{p}} = {\alpha _p} {\beta _p} {e^{j2\pi f{\tau _p}}}$.
Denote the channel between all BS antennas and the first port of the UE FA at time $t$ by
\begin{equation}
\begin{array}{l}
{{{\bf{ h}}_1}}(t) = {\left[ {{h_{1,1}}(t), \cdots ,{h_{{N_t},1}}(t)} \right]^T} = {{{\bf{ A}}}}{{\bf{C}}_1}(t)
\end{array},
\!\label{TraditionalChannelAllAntenna}
\end{equation}
where ${h_{n,1}(t)}$ is the channel between the $n$-th BS antenna and the first port of the UE FA:
\begin{equation}
\begin{array}{l}
{h_{n,1}(t)} = \sum\limits_{p = 1}^{P+1} {} {c_{p}}{e^{j2\pi \omega _p t}} {e^{\frac{{j2\pi {{({\bf{r}}_p^{\rm{tx}})}^T}{{\bf{d}}_{n}^{\rm{tx}}}}}{\lambda }}}
\end{array}.
\!\label{Channel-t}
\end{equation}
Without loss of generality, we select ${{{\bf{ h}}_1}}(t)$ as a reference static channel.
For the channel ${{{\bf{ h}}_1}}(t+\Delta t)$ at time $t + \Delta t$, we slide the liquid to the $m(\Delta t)$-th port, which makes the channel ${{{\bf{ h}}_{m(\Delta t)}}}(t+ \Delta t)$ close to the reference channel ${{{\bf{h}}_1}}(t)$.

\section{The moving port prediction method}\label{sec:MPMP prediction method}

This section will introduce our proposed MPMP prediction method.
By sliding the liquid to the optimal port, we may transform the time-varying channel to a static channel.
In this case, the BS estimates the channel parameters, e.g., angles, Doppler, and channel gains, and transmits them to the UE. 
After that, at the UE side, we construct an optimization problem to obtain the optimal port.

\subsection{The channel parameters estimation}

We will adopt the two-dimensional (2-D) MP method to estimate the angles, Doppler, and channel gains.
Denote the number of time samples used to estimate parameters by $N_s$, which is even.

Before introducing the 2-D MP matrix, we divide all time samples into two groups, i.e., the first and last half time samples.
We slide the FA at the first-half time samples and the last-half time samples to the ${\Delta _1}$-th and the ${\Delta _2}$-th port, respectively, where $\Delta _1 \ne \Delta _2$.
It is because we need to estimate the AOD and EOA by sliding the liquid to the different ports at different samples.
Define ${\chi _{\theta ,p}} = {{\cos {\theta _p}{d_v}}}$ and ${\kappa _{\theta ,\phi,p}} = {c_{p,m}} {e^{j\frac{{2\pi }}{\lambda }\sin {\theta _p}\sin {\phi _p}{d_h}}} $.

For the first half samples, we define a one-dimensional (1-D) MP matrix ${{\bf{G}}_{{n_t},{\Delta _1}}} \in {{\mathbb{C}}^{L \times (\frac{{{N_s}}}{2} - L + 1)}}$. 
For notational simplicity, we define $\mu_1 = \frac{{{N_s}}}{2} - L + 1$, where $L$ is the pencil size on the first half samples.
The matrix ${{\bf{G}}_{{n_t},{\Delta _1}}}$ is expressed as
\begin{equation}
\begin{array}{l}
{{\bf{G}}_{{n_t},{\Delta _1}}} = \left[ \begin{array}{l}
{h_{{n_t},{\Delta _1}}}(T), \ \ \cdots ,{h_{{n_t},{\Delta _1}}}({\mu_1}T)\\
\ \ \ \ \ \vdots \ \ \ \ \ \ \ \ \ \ddots  \ \ \ \ \ \ \ \ \ \vdots \\
{h_{{n_t},{\Delta _1}}}(LT), \cdots ,{h_{{n_t},{\Delta _1}}}(\frac{{{N_s}}}{2}T)
\end{array} \right]
\end{array},
\!\label{MP-FirstHalf}
\end{equation}
where ${P+1}<L<(\frac{{{N_s}}}{2} - {P} + 2)$, and $T$ is the sampling interval.
Then, we select the first column of the BS antennas, and generate the 2-D MP matrix ${{\bf{G}}_{\Delta _1}} \in {{\mathbb{C}}^{RL \times {\mu_1} {\mu_2}}}$ by
\begin{equation}
\begin{array}{l}
{{\bf{G}}_{\Delta _1}} = \left[ \begin{array}{l}
{{\bf{G}}_{1,{\Delta _1}}}, \ \cdots ,{{\bf{G}}_{\mu_2,{\Delta _1}}}\\
\ \ \ \vdots \ \ \ \ \ \ddots \ \ \ \ \ \ \vdots \\
{{\bf{G}}_{R,{\Delta _1}}}, \cdots ,{{\bf{G}}_{N_v,{\Delta _1}}}
\end{array} \right]
\end{array},
\!\label{2DMP-FirstHalf}
\end{equation}
where $R$ is the pencil size on the first column of the BS antennas, and $\mu _2 = N_v - R+1$.
By adopting the estimation algorithm in \cite{Li23TWC}, we estimate the Doppler ${{\omega} _p}$, the EOD parameter ${{{\chi }_{\theta ,p}}}$, and the parameter ${{c} _{p,{\Delta _1}}}$ of the $p$-th path as ${\hat{\omega} _p}$, ${{\hat{\chi }_{\theta ,p}}}$, and ${\hat {c} _{p,{\Delta _1}}}$.
Therefore, the EOD is estimated by ${\hat {\theta }_p} = \arccos (\frac{{{\hat {\chi }_{\theta ,p}}}}{{{d_v}}})$.

Next, following the procedure between Eq. (\ref{MP-FirstHalf}) and Eq. (\ref{2DMP-FirstHalf}), we also generate the second 2-D MP matrix ${{\bf{\tilde G}}_{\Delta _1}} \in {{\mathbb{C}}^{RL \times {\mu_1} {\mu_2}}}$ by updating Eq. (\ref{2DMP-FirstHalf}) with the second-column BS antennas.
Then, we estimate a parameter related to the path gain, EOD and EOA of the $p$-th path as ${\hat {\kappa} _{\theta ,\phi ,p}}$.

Similarly, with the last half samples, we generate a new 1-D MP matrix ${{\bf{G}}_{{n_t},{\Delta _2}}} \in {{\mathbb{C}}^{L \times (\frac{{{N_s}}}{2} - L + 1)}}$ by 
\begin{equation}
\!\!\!\!\begin{array}{l}
\!\!\!\!\!{{\bf{G}}_{{n_t},{\Delta _2}}} \!\!=\!\! \left[\!\! \begin{array}{l}
{h_{{n_t},{\Delta _2}}}((\frac{{{N_s}}}{2}\!+\!1)T), \ \cdots ,{h_{{n_t},{\Delta _2}}}((\frac{{{N_s}}}{2}\!+\!{\mu_1})T)\\
\ \ \ \ \ \ \ \ \ \ \ \vdots \ \ \ \ \ \ \ \ \ \ \ \ \ddots  \ \ \ \ \ \ \ \ \ \ \ \ \vdots \\
{h_{{n_t},{\Delta _2}}}((\frac{{{N_s}}}{2}\!+\!L)T), \cdots ,\ \ \ \ {h_{{n_t},{\Delta _2}}}({N_s}T)
\!\!\end{array}\!\!\!\! \right]\!\!\!
\end{array}.
\!\!\!\!\label{MP-LastHalf}
\end{equation}
Then, by exploiting the channel between the first-column BS antennas and the UE, we extend the 1-D to the 2-D case as:
\begin{equation}
\begin{array}{l}
{{\bf{G}}_{\Delta _2}} = \left[ \begin{array}{l}
{{\bf{G}}_{1,{\Delta _2}}}, \ \cdots ,{{\bf{G}}_{\mu_2,{\Delta _2}}}\\
\ \ \ \vdots \ \ \ \ \ \ddots \ \ \ \ \ \ \vdots \\
{{\bf{G}}_{R,{\Delta _2}}}, \cdots ,{{\bf{G}}_{N_v,{\Delta _2}}}
\end{array} \right]
\end{array}.
\!\label{2DMP-LastHalf}
\end{equation}
Define ${\varpi _{p,m}} = {c_{p,m}} {e^{j2\pi \omega _p {\frac{{{N_s}}}{2}}}} $.
We obtain the estimations of the Doppler ${{\omega} _p}$, the EOD parameter ${\chi _{\theta ,p}}$, parameter ${\varpi _{p,{\Delta _2}}}$, and the number of paths as: ${\hat{\tilde \omega} _p}$, ${\hat {\tilde \chi} _{\theta ,p}}$, ${\hat {\varpi }_{p,{\Delta _2}}}$, and $\hat {\cal P}$.

To determine the estimations of AOD and EOA, we pair the estimations of Doppler ${\hat{\bm{\omega}}}$ and ${\hat{\tilde{\bm{\omega}}}}$ by the pairing algorithm in \cite{Li23TWC}.
We denote the corresponding estimations of parameters after pairing by several $(P+1)\times 1$ vectors: ${{\hat{\bm{{\chi}}}_{\theta }}}$, ${\hat {\bm {c}} _{{\Delta _1}}}$, ${\hat {\bm {\kappa}} _{\theta ,\phi }}$ and ${\hat {\bm {\varpi }}_{{\Delta _2}}}$.
The $p$-th entries of ${\hat{\bm{\omega}}}$, ${{\hat{\bm{{\chi}}}_{\theta}}}$, ${\hat {\bm {c}} _{{\Delta _1}}}$, ${\hat {\bm {\kappa}} _{\theta ,\phi }}$ and ${\hat {\bm {\varpi }}_{{\Delta _2}}}$ are ${\hat{{\omega}}_p}$, ${\hat \chi _{p}}$, ${\hat { {c}} _{p,{\Delta _1}}}$, ${\hat \kappa _{p}}$ and ${\hat {{\varpi }}_{p,{\Delta _2}}}$.
The AOD and EOA of the $p$-th path is estimated as 
\begin{equation}
\begin{array}{l}
{\hat {\phi} _p} = \arcsin (\frac{{\lambda {d_v}\angle (\frac{{{{\hat \kappa }_p}}}{{{{\hat \theta }_p}}})}}{{2\pi {d_h}\sqrt {d_v^2 - \hat \chi _p^2} }})
\end{array},
\!\label{AOD-est}
\end{equation}
\begin{equation}
\begin{array}{l}
{\hat {\theta} _p^{\rm{rx}}} = \arccos (\frac{{\angle (\frac{{{{\hat \varpi }_{p,{\Delta _2}}}}}{{{{\hat c}_{p,{\Delta _1}}}}}{e^{ - j{N_s}\pi {{\hat \omega }_p}}})}}{{2\pi d_v^{{\rm{rx}}}({\Delta _2} - {\Delta _1})}})
\end{array}.
\!\label{EOA-est}
\end{equation}

With the estimations of EOD ${\hat \theta _p}$, AOD ${{\hat \phi }_p}$, EOA ${\hat {\theta} _p^{\rm{rx}}}$, Doppler ${\hat{\omega} _p}$, and channel gains ${\hat { {c}} _p}$ of all paths, we may reconstruct the channel between all BS antennas and the $m$-th port of the UE FA at time $t$ as ${{{\bf{\hat { h}}}}_m}(t)$ in Eq. (\ref{ChannelAllAntenna}).

Based on the reconstructed channel, we will design a moving port prediction method.
To determine the optimal port, in Sec. \ref{sec:relationship}, we construct the optimization problem based on the FA channel ${{{\bf{ h}}}_{m}}(t)$ in Eq. (\ref{ChannelAllAntenna}) and the reference channel ${{{\bf{ h}}_1}}(t)$ in Eq. (\ref{TraditionalChannelAllAntenna}).

\subsection{The optimization problem}\label{sec:relationship}

Denote the FA channel of the $m( \Delta t)$-th port between the BS antennas and the UE at time $t + \Delta t$ by
\begin{equation}
\!\!\!\!\!\begin{array}{l}
{{{\bf{ h}}_{m(\Delta t)}}}(t+ \Delta t) \!=\! {{{\bf{ A}}}}{{\bf{ C}}_{m(\Delta t)}}(t + \Delta t).
\end{array}
\!\!\!\label{FluidChannelAllAntenna-t-dt}
\end{equation}
According to Eq. (\ref{TraditionalChannelAllAntenna}), we define the error vector between ${{{\bf{h}}}_{m(\Delta t)}}(t+ \Delta t)$ and ${{{\bf{h}}_1}}(t)$ as:
\begin{equation}
\begin{array}{l}
\!\!\!\!\!\!\!{\bm{\varepsilon }} = {\left[ {{\varepsilon}_1,\! \cdots \!,{{\varepsilon}_{N_t}}} \right]^T} = {{{\bf{ h}}_{m(\Delta t)}}}(t + \Delta t) - {{{\bf{ h}}_1}}(t)\\
={{{\bf{ A}}}}\!(\!{{\bf{ C}}_{m}}\!(t \!+ \!\Delta t) \!-\! {{\bf{C}}_1}(t))
=\! \sum\limits_{p = 1}^{P+1} {} \!{c_{p}}{e^{j2\pi \omega _p t}}{\bf{a}}({\theta _p},\!{\phi _p})\\
 ({e^{j2\pi {\omega _p}\Delta t}}\!{e^{j\!\frac{{2\pi }}{\lambda }\cos {{\theta} _p^{\rm{rx}}}d_v^{{\rm{rx}}}(m(\Delta t) -\! 1)}} \!-\! 1),
\end{array}
\!\!\label{AllErrorFluidAllAntenna}
\end{equation}
where ${\varepsilon}_n$, $n=1,\cdots, N_t$, is the error between ${h_{n,{m(\Delta t)}}}(t+\Delta t)$ and ${h_{n,1}(t)}$:
\begin{equation}
\!\!\!\!\!\!\!\begin{array}{l}
{\varepsilon}_n = \sum\limits_{p = 1}^{P+1} {} {c_{p}}{e^{j\frac{{2\pi }}{\lambda }\sin {\theta _p}\sin {\phi _p}{d_h}(n_h - 1)}}{e^{j\frac{{2\pi }}{\lambda }{d_v}\cos {\theta _p}({n_v} - 1)}} \\
\ \ \ \ \ \ \ \ \ {e^{j2\pi {\omega _p}t}}({e^{j2\pi {\omega _p}\Delta t}}{e^{j\frac{{2\pi }}{\lambda }{d_v^{\rm{rx}}}\cos {{\theta} _p^{\rm{rx}}}(m(\Delta t) - 1)}} - 1)\\
\ \ \ \ = 2\sum\limits_{p = 1}^{P+1} {} {\alpha _p}{\beta _p}\sin {\varsigma _p}{e^{j(\frac{\pi }{2} + {\delta _p} + {\varsigma _p})}},
\end{array}
\!\!\!\!\!\!\!\!\!\!\!\!\!\!\!\!\!\label{ErrorFluidAllAntennaIni}
\end{equation}
with 
\begin{equation}
\begin{array}{l}
{\delta _p} = 2\pi f{\tau _p} + 2\pi {\omega _p}t + \frac{{2\pi }}{\lambda }\cos {\theta _p}{d_v}({n_v} - 1) \\
\ \ \ + \frac{{2\pi }}{\lambda }\sin {\theta _p}\sin {\phi _p}{d_h}({n_h} - 1),
\end{array}
\!\!\!\!\label{Parameter1}
\end{equation}
\begin{equation}
\begin{array}{l}
\!\!\!\!\!\!\!\!\!\!\!\!\!\!\!\!\!\!\!\!{\varsigma _p} = \pi {\omega _p}\Delta t + \frac{{\pi \cos {{\theta} _p^{\rm{rx}}}(m(\Delta t) - 1)d_v^{{\rm{rx}}}}}{\lambda }.
\end{array}
\!\!\!\!\label{Parameter2}
\end{equation}
We construct an optimization problem:
\begin{equation}
\begin{array}{l}
{m(\Delta t)} = \mathop {\arg \min }\limits_{{m(\Delta t)}} (\sum\limits_{{n} = 1}^{{N_t}} {\left\| {{\varepsilon _{{n}}}} \right\|_2^2}).
\end{array}
\!\label{OptimizationProblem}
\end{equation}
By sliding the liquid to the ${m(\Delta t)}$-th port, we may transform the time-varying channel ${{{\bf{h}}}_{1}}(t+ \Delta t)$ in Eq. (\ref{FluidChannelAllAntenna-t-dt}) to ${{{\bf{h}}}_{m(\Delta t)}}(t+ \Delta t)$, which is close to the static channel ${{{\bf{h}}_1}}(t)$ in Eq. (\ref{TraditionalChannelAllAntenna}).
Based on Eq. (\ref{ErrorFluidAllAntennaIni}), we compute ${\left\| {{\varepsilon _{{n}}}} \right\|_2^2}$ as:
\begin{equation}
\begin{array}{l}
\left\| {{\varepsilon _n}} \right\|_2^2 = 4{\left| {\sum\limits_{p = 1}^{P+1} {} {\alpha _p}{\beta _p}\sin {\varsigma _p}{e^{j(\frac{\pi }{2} + {\delta _p} + {\varsigma _p})}}} \right|^2}.
\end{array}
\!\label{Error-n-2}
\end{equation}

Next, we categorize the channels into two cases: a LoS channel and a multi-path channel, and design port selection methods in different cases.

\subsection{The port selection method for a LoS channel}\label{PortSelectionLoS}

The parameter ${\left\| {{\varepsilon _{{n}}}} \right\|_2^2}$ in Eq. (\ref{Error-n-2}) is calculated by:
\begin{equation}
\begin{array}{l}
\left\| {{\varepsilon _n}} \right\|_2^2 = 4{\left|{\alpha ^{{\rm{LoS}}}} {{\beta ^{{\rm{LoS}}}}\sin {\varsigma ^{{\rm{LoS}}}}{e^{j(\frac{\pi }{2} + {\delta ^{{\rm{LoS}}}} + {\varsigma ^{{\rm{LoS}}}})}}} \right|^2} \\
\ \ \ \ \ \ \ \ = \!\frac{{4{K_R}}}{{1 + {K_R}}}{\sin ^2}{\varsigma ^{{\rm{LoS}}}},
\end{array}
\!\label{Error-n-ALoS}
\end{equation}
where ${\delta ^{{\rm{LoS}}}}$ and ${\varsigma ^{{\rm{LoS}}}}$ are obtained by updating all the parameters of ${\delta _p}$ and ${\varsigma _p}$ in Eq. (\ref{Parameter1}) and Eq. (\ref{Parameter2}) with some LoS-path parameters, e.g., ${\omega ^{\rm{LoS}}}$, ${\tau ^{\rm{LoS}}}$, ${\theta _{{\rm{EoD}}}^{{\rm{LoS}}}}$, ${\phi _{{\rm{AoD}}}^{{\rm{LoS}}}}$, and ${\theta _{{\rm{EoA}}}^{{\rm{LoS}}}}$.
The optimization problem in Eq. (\ref{OptimizationProblem}) is transformed to
\begin{equation}
\begin{array}{l}
{m(\Delta t)} = \mathop {\arg \min }\limits_{{m(\Delta t)}} ({\sin ^2}{\varsigma ^{{\rm{LoS}}}})
\end{array}.
\!\label{OptimizationProblem-LoS}
\end{equation}
We make a technical assumption that ${\theta _{\rm{EOA}}^{{\rm{LoS}}}} \ne \frac{\pi }{2}$.
It is reasonable when the BS is higher than the UE.
Therefore, $\cos {\theta _{\rm{EOA}}^{{\rm{LoS}}}} \ne 0$.
Let ${\sin ^2}{\varsigma ^{{\rm{LoS}}}} = 0$, we obtain:
\begin{equation}
\begin{array}{l}
{m(\Delta t)} = \left[ {\lambda \frac{{\bmod (\pi \omega ^{{\rm{LoS}}}\Delta t,\pi )}}{{\pi {d_v^{\rm{rx}}}\cos \theta _{\rm{EOA}}^{{\rm{LoS}}}}}} \right] + 1\\
\ \ \ \ \ \ \ \ \ =\left[ {\frac{{\rho ( - {\omega ^{{\rm{LoS}}}}\Delta t +  k(\Delta t))}}{{\cos {\theta _{\rm{EOA}}^{{\rm{LoS}}}}}}} \right] + 1,
\end{array}
\!\label{Port-LoS}
\end{equation}
where $k(\Delta t) \in {\bf{Z}}$ is used to ensure ${m(\Delta t)}>0$.
The selected global optimal port should be periodic with a period of ${T_{\rm{LoS}}}$:
\begin{equation}
\begin{array}{l}
{T_{{\rm{LoS}}}} \approx \left[ \frac{{\rho}}{{\cos \theta _{{\rm{EOA}}}^{{\rm{LoS}}}}} \right].
\end{array}
\!\label{LoSTermsPeriod-1}
\end{equation}
If FA contains at least a period, e.g., the FA length satisfies $W\lambda \ge {T_{{\rm{LoS}}}}d_v^{{\rm{rx}}}$, we obtain the global optimal port shown in Eq. (\ref{Port-LoS}).
Due to $1 \le {m(\Delta t)} \le M$, the range of $k(\Delta t)$ is
\begin{equation}
\begin{array}{l}
{\bar A} \le k(\Delta t) \le {\bar B},
\end{array}
\!\label{Port-LoSk}
\end{equation}
where 
\begin{equation}
\begin{array}{l}
{\bar A} = \min({{{\omega ^{{\rm{LoS}}}}\Delta t}}, {{W\cos {\theta _{\rm{EOA}}^{{\rm{LoS}}}} + {\omega ^{{\rm{LoS}}}}\Delta t}}),
\end{array}
\!\label{Port-LoSkA}
\end{equation}
\begin{equation}
\begin{array}{l}
{\bar B} = \max({{{\omega ^{{\rm{LoS}}}}\Delta t}}, {{W\cos {\theta _{\rm{EOA}}^{{\rm{LoS}}}} + {\omega ^{{\rm{LoS}}}}\Delta t}}).
\end{array}
\!\label{Port-LoSkB}
\end{equation}
However, in case $W\lambda < {T_{{\rm{LoS}}}}d_v^{{\rm{rx}}}$, we let $x=(m(\Delta t) - 1)d_v^{{\rm{rx}}}$ and $g(x) = {\sin ^2}(\pi {\omega ^{{\rm{LoS}}}}\Delta t + \frac{{\pi \cos \theta _{{\rm{EOA}}}^{{\rm{LoS}}}}}{\lambda }x)$.
The optimization problem in Eq. (\ref{OptimizationProblem-LoS}) now involves finding the minimum value of $g(x)$ over the interval $[0,W\lambda]$.
Evidently, $g(x)$ has, at most, one extreme point denoted by $x_0$. 
The minimum value of $g(x)$ is determined by ${(g(x))_{\min }} = \min (g(0),g(W\lambda ),g({x_0}))$.
Thus, the global optimal port is:
\begin{equation}
\begin{array}{l}
m(\Delta t) \approx \left[ {\frac{{\{ x|{{(g(x))}_{\min }}\} }}{{d_v^{{\rm{rx}}}}}} \right] + 1.
\end{array}
\!\label{Port-LoS-best-min}
\end{equation}

\subsection{The port selection method for a multi-path channel}

According to Eq. (\ref{Error-n-2}), ${\left\| {{\varepsilon _n}} \right\|_2^2}$ is firstly calculated by:
\begin{equation}
\!\!\begin{array}{l}
\!\!\!\!\!\!\!\!\left\| {{\varepsilon _n}} \right\|_2^2 = 4\!\!\sum\limits_{p = 1}^{P+1} {}{\alpha _p^2} {\beta _p^2}{\sin ^2}{\varsigma _p} \\
\!+ 8\!\!\sum\limits_{p = 1}^{P} {}\!\! \sum\limits_{q = p + 1}^{P+1} {}\!\! {\alpha _p}{\alpha _q}{\beta _p}{\beta _q}\sin {\varsigma _p}\sin {\varsigma _q}\cos ({\delta _p} \!+\! {\varsigma _p}\! - \!{\delta _q}\! -\! {\varsigma _q}).
\end{array}
\!\!\!\!\!\!\!\!\!\!\!\label{ErrorFluidAllAntenna-Derivation}
\end{equation}
Then, we calculate $\sum\limits_{{n} = 1}^{{N_t}} {\left\| {{\varepsilon _{{n}}}} \right\|_2^2}$ in Eq. (\ref{OptimizationProblem}) as:
\begin{equation}
\begin{array}{l}
\sum\limits_{n = 1}^{{N_t}} {} \left\| {{\varepsilon _n}} \right\|_2^2 \!= \!4{N_t}\sum\limits_{p = 1}^{P+1} {}{\alpha _p^2} {\beta _p^2}{\sin ^2}{\varsigma _p} \\
\ \ \ \ \ \ \ \ \ \ \ \ \ + 8\!\sum\limits_{p = 1}^{P} {} \!\!\sum\limits_{q = p + 1}^{P+1} {} \!{\alpha _p}{\alpha _q}{\beta _p}{\beta _q}\sin {\varsigma _p}\sin {\varsigma _q} {\Upsilon _{p,q}},
\end{array}
\!\label{ErrorPeriod-all}
\end{equation}
where
\begin{equation}
\!\!\!\!\!\!\begin{array}{l}
 {\Upsilon _{p,q}}=
\sum\limits_{{n_v} = 1}^{{N_v}} {} \sum\limits_{{n_h} = 1}^{{N_h}} {} \cos ({a_{p,q}}{n_v} + {b_{p,q}}{n_h} + {\xi _{p,q}})\\
\ \ \ \ \mathop  = \limits^{(a)} \!\sin\!\frac{{{N_h}}{b_{p,q}}}{2}\!\csc\frac{{{b_{p,q}}}}{2}\!\!\sum\limits_{{n_v} = 0}^{{N_v} - 1} {} \!\!\cos({\xi _{p,q}} \!+\! \frac{{{N_h} - 1}}{2}{b_{p,q}}\! +\! {n_v}{a_{p,q}})\\
\ \ \ \ \mathop  = \limits^{(b)} \sin\frac{{{N_h}{b_{p,q}}}}{2}\csc\frac{{{b_{p,q}}}}{2}\sin\frac{{{N_v}{a_{p,q}}}}{2}\csc\frac{{{a_{p,q}}}}{2}\\
\ \ \ \ \ \ \ \ \ \cos({\xi _{p,q}} + \frac{{{N_h} - 1}}{2}{b_{p,q}} + \frac{{{N_v} - 1}}{2}{a_{p,q}}),
\end{array}
\!\!\!\!\!\!\label{ErrorPeriod-all-sub}
\end{equation}
with ${a_{p,q}} = \frac{{2\pi }}{\lambda }(\cos {\theta _p} - \cos {\theta _q}){d_v}$, ${b_{p,q}} = \frac{{2\pi }}{\lambda }(\sin {\theta _p}\sin {\phi _p} - \sin {\theta _q}\sin {\phi _q}){d_h}$, and 
\begin{equation}
\begin{array}{l}
{\xi _{p,q}} = 2\pi f({\tau _p} - {\tau _q}) + \pi (2t + \Delta t)({\omega _p} - {\omega _q}) \\
\ \ \ \ \ + \frac{{\pi (m - 1)d_v^{{\rm{rx}}}}}{\lambda }(\cos {{\theta} _p^{\rm{rx}}} - \cos {{\theta} _q^{\rm{rx}}}).
\end{array}
\!\label{ErrorPeriod-all-parameterxi}
\end{equation}
In Eq. (\ref{ErrorPeriod-all-sub}), $(a)$ and $(b)$ are achieved by the sum formula of the trigonometric function in \cite{TableIntegral}
\begin{equation}
\!\!\!\begin{array}{l}
\sum\limits_{{\bar k} = 1}^{\bar n} {} \cos ({\bar x} + ({\bar k} - 1){\bar y}) = \cos ({\bar x} + \frac{{{\bar n} - 1}}{2}{\bar y})\sin\frac{{{\bar n}{\bar y}}}{2}\csc \frac{{\bar y}}{2}.
\end{array}
\!\!\!\label{SumFormula1}
\end{equation}
For the $p$-th path, searching the optimal port that satisfies $\min (\sum\limits_{{n} = 1}^{{N_t}} {\left\| {{\varepsilon _{{n}}}} \right\|_2^2})$ in Eq. (\ref{ErrorPeriod-all}), equates to finding the port that satisfies $\sin {\varsigma _p} = 0$.
Let $\sin {\varsigma _p} = 0$, and we obtain the optimal port of the $p$-th path:
\begin{equation}
\begin{array}{l}
{m_p}(\Delta t) \approx \left[ \frac{{\lambda ({k_p}(\Delta t) - {\omega _p}\Delta t)}}{{d_v^{{\rm{rx}}}\cos {{\theta} _p^{\rm{rx}}}}} \right] + 1,
\end{array}
\!\label{BestPorts-EachPaths}
\end{equation}
where $k_p(\Delta t) \in {\bf{Z}}$ ensures ${m_p}(\Delta t) >0$.
In Eq. (\ref{ErrorPeriod-all}), the non-cross term of LoS and NLoS paths ${\alpha _p^2}{\beta _p^2}{\sin ^2}{\varsigma _p}$ and the cross term of two different paths ${\alpha _p}{\alpha _q}{\beta _p}{\beta _q}\sin {\varsigma _p}\sin {\varsigma _q}{\Upsilon _{p,q}}$ should be periodic functions with a period of ${T_{p}}$ and ${T_{p,q}}$, respectively:
\begin{equation}
\begin{array}{l}
{T_{p}} = {T_{p,q}} \approx \left[ {\frac{\lambda }{{d_v^{{\rm{rx}}}\cos {{\theta} _p^{\rm{rx}}}}}} \right].
\end{array}
\!\label{CrossTermsPeriod-1}
\end{equation}
Therefore, $\sum\limits_{{n} = 1}^{{N_t}} {\left\| {{\varepsilon _{{n}}}} \right\|_2^2}$ should be a periodic function with a period of $T_{{\varepsilon}}$:
\begin{equation}
\begin{array}{l}
{T_{\varepsilon }} \approx lcm\left( {{T_{1}}, \cdots ,{T_{{p}}}, \cdots {{T_{P}}}} \right),
\end{array}
\!\label{ErrorPeriod-n}
\end{equation}
which indicates that the period ${T_{\varepsilon }}$ is independent of the number of BS antennas.
If the FA contains the period ${T_{\varepsilon }}$, the smallest FA length satisfies 
\begin{equation}
\begin{array}{l}
{{W\lambda }} \ge L_{{\varepsilon}} = {T_{\varepsilon }} {d_v^{\rm{rx}}}.
\end{array}
\!\label{ErrorPeriod-FluidLength}
\end{equation}

Define a function $f(x)$, $0 \le x \le (M-1){d_v^{\rm{rx}}}$, such that $x = (m(\Delta t)-1){d_v^{\rm{rx}}}$ and $f(x) =\sum\limits_{{n} = 1}^{{N_t}} {\left\| {{\varepsilon _{{n}}}} \right\|_2^2}$.
Evidently, $f(x)$ is also a periodic function with a period of $L_{{\varepsilon}}$.

Now, we transform the problem of searching the minimum value of $\sum\limits_{{n} = 1}^{{N_t}} {\left\| {{\varepsilon _{{n}}}} \right\|_2^2}$ to searching the minimum value of $f(x)$.
Define the minimum value of $f(x)$ as $({f}(x))_{\min }$.The global optimal port $m(\Delta t)$ is determined by
\begin{equation}
\begin{array}{l}
{m(\Delta t)} \approx \left[ \frac{\left\{ {\left[ x \right]|({f}(x))_{\min }} \right\}}{{d_v^{\rm{rx}}}} \right] + 1.
\end{array}
\!\label{mn-multiple}
\end{equation}

Next, we will compute the $({f}(x))_{\min }$.
Compute the derivative of $f(x)$ as $\frac{{\partial f(x)}}{{\partial x}}$.
Given $x \in [0,{{W\lambda }}]$, if $\frac{{\partial f(x)}}{{\partial x}}>0$ or $\frac{{\partial f(x)}}{{\partial x}}<0$, $f(x)$ is a monotonically increasing function or a monotonically decreasing function.
$({f}(x))_{\min }$ is calculated by
\begin{equation}
\begin{array}{l}
({f}(x))_{\min } = \min \left( f(0),f({{W\lambda }}) \right)
\end{array}.
\!\label{Minfx-monotone}
\end{equation}
If $f(x)$ is not a monotonic function, we let $\frac{{\partial f(x)}}{{\partial x}} = 0$, and obtain a set ${\mathfrak{X}}$ including all extreme points:
\begin{equation}
\begin{array}{l}
\!\!\!\!\!{\mathfrak{X}} \!=\! \left\{ {x|\frac{{\partial f(x)}}{{\partial x}} \!=\! 0, \frac{{{\partial ^2}f(x)}}{{\partial {x^2}}} \!>\! 0, 0 \le x \le \min ({L_\varepsilon },{{W\lambda }})} \right\}\\
\!\!\ = \left\{ {{x_1}, \cdots ,{x_{{M_x}}}} \right\},
\end{array}
\!\label{ExtremePoints}
\end{equation}
where $M_x$ is the number of extreme points.
If ${L_\varepsilon } < {{W\lambda }}$, ${\mathfrak{X}}$ includes all extreme points during a period.
As ${L_\varepsilon } \ge {{W\lambda }}$, the set ${\mathfrak{X}}$ contains all extreme points during an interval of $[0,{{W\lambda }}]$.
Then, the minimum value is 
\begin{equation}
\!\!\!\!\begin{array}{l}
({f}(x))_{\min } = \min \left( {\left\{ {f(x)|x \in \left\{ 0,{{W\lambda }} \right\} \cup {\mathfrak X}} \right\}} \right).
\end{array}
\!\label{Minfx-2}
\end{equation}

Up to now, at time $t+\Delta t$, for a time-varying channel ${{{\bf{ h}}_1}}(t+\Delta t)$ between the BS antenna and the first port of the UE FA, we slide the liquid to the $m(\Delta t)$-th port and transform the channel ${{{\bf{ h}}_1}}(t+\Delta t)$ to ${{{\bf{ h}}_{m(\Delta t)}}}(t+ \Delta t)$, which is close to the static channel ${{{\bf{ h}}_1}}(t)$.

The detailed design process is illustrated in Algorithm \ref{algorithm_MPMP}.
We may notice that the FA length $W\lambda$ and the number of ports $M$ affect the accuracy of the selected port.
However, when the length is large enough, e.g., ${{W\lambda }} \ge {L_\varepsilon }$ in a multi-path channel and $W\lambda \ge {T_{{\rm{LoS}}}}d_v^{{\rm{rx}}}$ in a LoS channel, a further extended length will not benefit the performance, which will be analyzed in the next section.
\begin{algorithm}[htb]
\renewcommand{\algorithmicrequire}{\textbf{Input:}}
\renewcommand{\algorithmicensure}{\textbf{Output:}}
\caption{The matrix pencil-based moving port prediction scheme.}
\label{algorithm_MPMP}
\begin{algorithmic}[1]
\REQUIRE $N_s$, ${{{\bf{ h}}}_{m}}(t)$, ${\Delta _1}$, ${\Delta _2}$;
\STATE Estimate the channel ${{{\bf{\hat { h}}}}_m}(t)$ and channel parameters, e.g., the Doppler ${\hat{\bm{\omega}}}$, EOD ${\hat {\theta }_p}$, AOD ${\hat {\phi} _p}$, EOA ${\hat {\theta} _p^{\rm{rx}}}$, path gain ${\hat { {c}} _p}$, and the number of paths $\hat {\cal P}$;
\STATE Construct the optimization problem as shown in Eq. (\ref{OptimizationProblem}) and Eq. (\ref{Error-n-2});
\IF{${\hat {\cal P}} = 1$}
\IF{$W\lambda \ge {T_{{\rm{LoS}}}}d_v^{{\rm{rx}}}$}
\STATE Select the optimal port by Eq. (\ref{Port-LoS});
\ELSE
\STATE Select the optimal port by Eq. (\ref{Port-LoS-best-min});
\ENDIF
\ELSE
\STATE Select the optimal port of the multi-path channel by Eq. (\ref{mn-multiple});
\ENDIF
\ENSURE ${m(\Delta t)}$
\end{algorithmic}
\end{algorithm}

\section{Performance Analysis}\label{sec:Performance analysis}

In this section, we start the performance analysis of our proposed matrix pencil-based moving port prediction method by analyzing the step phenomenon in sliding liquid.
Then, we prove the asymptotic prediction MSE for the LoS channel.
Finally, we extend the LoS channel to a multi-path channel, and derive the upper and lower asymptotic bounds of the prediction MSE.

Before the analysis, we introduce a technical assumption.
\begin{assumption}\label{Assumption-Mild}
The observation sample satisfies
\begin{equation}
{\check{\bf{h}}_m}(t) = {{\bf{h}}_m}(t) + {\bf{n}},
\!\!\!\label{observation samples}
\end{equation}
where ${\bf{n}} \in {{\mathbb{C}}^{{N_t} \times 1}}$ is the independent identically distributed (i.i.d.) Gaussian white noise with zero mean and element-wise variance ${\sigma ^2}$:
\begin{equation}
{\bf{n}} = {[{n_1}, \cdots ,{n_{{N_t}}}]^T}.
\!\!\!\label{Noise}
\end{equation}
As $N_t \to \infty $, the variance  ${\sigma ^2}$ converges to zero, such that:
\begin{equation}
\mathop {\lim }\limits_{N_t \to \infty } \frac{{\left\| {{{{\bf{\check h}}}_m}(t) - {{\bf{h}}_m}(t)} \right\|_2^2}}{{\left\| {{{\bf{h}}_m}(t)} \right\|_2^2}} = 0.
\!\!\label{Samples_accurate}
\end{equation}
\end{assumption}
Remarks: This technical assumption means the normalized channel sample error converges to zero when the number of BS antennas increases. The condition of Eq. (\ref{Samples_accurate}) can be achieved even in the multi-user multi-cell scenario with pilot contamination, by some non-linear signal processing technologies \cite{Yin16TSP}. 

Since our proposed MPMP prediction method is based on the estimations of parameters, we first study the estimation accuracy of parameters in Proposition \ref{proposition1}.
\begin{Proposition}\label{proposition1} 
Under Assumption \ref{Assumption-Mild}, if three samples are available, the asymptotic performance of the estimated parameters yields
\begin{equation}
\mathop {\lim }\limits_{N_t \to \infty } \left[{\hat{\phi }_p, \hat{\theta }_p, { {\hat {\theta} _p^{\rm{rx}}}},{\hat{\omega }_p},{{\hat{c}_{p}}}} \right] = \left[{{\phi }_p, {\theta }_p, {{\theta} _p^{\rm{rx}}}, {{\omega }_p},{c_{p}}} \right],
\!\!\!\!\label{CSI delay: parameters estimation}
\end{equation}
where $p = 1, \cdots, {P+1}$.
\end{Proposition}
\begin{proof} 
When the number of the BS antennas is large enough, this Proposition is a simplified version of Theorem 1 in \cite{Li23TWC} since the noise is i.i.d. Gaussian white noise among the antennas.
Based on Eq. (\ref{MP-FirstHalf}) and Eq. (\ref{2DMP-FirstHalf}), the 2-D MP matrix ${\bf{D}} \in {{\mathbb{C}}^{RL \times {\mu_1} {\mu_2}}}$, composed of the observation samples in Eq. (\ref{observation samples}), is calculated by ${\bf{D}} = {{\bf{G}}_{\Delta _1}} + {\bf{N}}$, where ${\bf{N}}$ is a noise matrix.
Given just two samples and enough BS antennas, the rank of ${{\bf{G}}_{\Delta _1}}$ satisfies $r({{\bf{G}}_{\Delta _1}})>{P+1}$, which means that the matrix ${{\bf{G}}_{\Delta _1}}$ contains the information of all paths.
Perform the correlation matrix of ${\bf{D}}$ as 
\begin{equation}
\begin{array}{l}
\mathop {\lim }\limits_{N_v \to \infty } {{\bm{R}}_{{{{\bf{D}}}}}} = E\{ {{{\bf{D}}}}{{{{\bf{D}}}}^H}\}= {{\bm{R}}_{{{\bf{G}}_2}}} + {\sigma ^2}{{\bf{I}}_{{N_v}{N_s}}},
\end{array}
\!\label{Correlation}
\end{equation}
where ${\sigma ^2}{{\bf{I}}_{{N_v}{N_s}}} = E\{ {{{\bf{N}}}}{{{{\bf{N}}}}^H}\}$ and $E\{ \cdot\}$ is the expectation over the BS antennas.
Then, calculate the SVD of ${\bf{D}}$ and estimate the channel parameters, e.g., EOD $\hat{\phi }_p$ and Doppler ${\hat{\omega }_p}$.
Likewise, generate two 2-D MP matrices, and estimate the AOD $\hat{\theta }_p$, EOA ${{\hat {\theta} _p^{\rm{rx}}}}$, and channel gain ${{\hat{c}_{p}}}$.
We may obtain that $\mathop {\lim }\limits_{N_t \to \infty } \left[{\hat{\phi }_p, \hat{\theta }_p, {{\hat \theta} _p^{\rm{rx}}}, {\hat{\omega }_p},{{\hat{c}_{p}}}} \right] = \left[{{\phi }_p, {\theta }_p, {{\theta} _p^{\rm{rx}}}, {{\omega }_p},{c_{p}}} \right]$, where $p = 1, \cdots, {P+1}$.
The detailed proof is omitted.
\end{proof}
Remarks: Proposition 1 indicates that the estimated parameters converge to error-free if the BS has enough antennas.
Based on Proposition 1, we also obtain that for an arbitrary CSI delay $\tau$, the estimation and prediction errors of the FA channel converge to error-free, i.e.,
\begin{equation}
\begin{array}{l}
\mathop {\lim }\limits_{{N_t},{\rho} \to \infty } \frac{\left\| {{{{\bf{\hat { h}}}_{m(\tau)}}}(t + \tau) \!-\! {{{\bf{ { h}}}_{m(\tau)}}}(t + \tau)} \right\|_2^2}{{\left\| {{{\bf{ { h}}}_{m(\tau)}}}(t + \tau) \right\|_2^2}} \!= 0.
\end{array}
\!\label{Correlation}
\end{equation}
Proposition 1 is also the prior basis of the following performance analysis.
Next, we will introduce the step performance of sliding liquid in Lemma \ref{Lemma1}.

\begin{lemma}\label{Lemma1}
Under Assumption \ref{Assumption-Mild}, for the LoS channel, if the fluid length satisfies $W\lambda > \frac{\lambda}{{\left| {\cos {\theta _{\rm{EOA}}^{{\rm{LoS}}}}} \right|}}$, the speed of sliding liquid satisfies
\begin{equation}
\begin{array}{l}
{v_f}(t_1) < {v_f}(t_2) \le \frac{{W\lambda }}{T},
\end{array}
\!\label{SpeedStep}
\end{equation}
where ${v_f}(t_1)$ and ${v_f}(t_2)$ are the sliding speeds at time $t_1$ and $t_2$.
\end{lemma}
\begin{proof}
The proof can be found in Appendix A.
\end{proof}
Remarks:
Lemma \ref{Lemma1} illustrates the step phenomenon of sliding liquid. 
Specifically, let ${m}(t_1)$ denote the selected port at time $t_1$ and ${\mathfrak n}T$ denote a maximum time interval that satisfies ${m}(t_1) +{v_f(t_1)}{\mathfrak n}{T} \le M<{m}(t_1) + {v_f(t_1)}({\mathfrak n}+1){T}$.
Over the interval of $[0, {\mathfrak n}{T}]$, the global optimal port is achieved by sliding the liquid at the speed of ${v_f(t_1)}$.
However, at time $({\mathfrak n}+1){T}$, sliding the liquid at the speed of ${v_f(t_1)}$ results in the selected port larger than $M$, which is not feasible.
Fortunately, as shown in Eq. (\ref{LoSTermsPeriod-1}), the global optimal port is periodic.
To select the global optimal port at time $({\mathfrak n}+1){T}$, we need to slide the liquid at the speed of ${v_f(t_2)}$ in the opposite direction.
The above sliding speed variation is the step phenomenon of sliding liquid.

Lemma \ref{Lemma1} also indicates the speed of sliding liquid is related to the wavelength $\lambda$ and the EOA ${\theta _{\rm{EOA}}^{{\rm{LoS}}}}$.
The sliding speed is smaller for a channel with a higher carrier frequency.
Lemma \ref{Lemma1} demonstrates the minimum FA length, e.g., $W\lambda > \frac{\lambda}{{\left| {\cos {\theta _{\rm{EOA}}^{{\rm{LoS}}}}} \right|}}$, includes at least one global optimal port.
When extending Lemma \ref{Lemma1} to a multi-path channel, if the FA length satisfies $W\lambda > {T_{\varepsilon }} {d_v^{\rm{rx}}}$ and the FA contains the global optimal port, the step phenomenon of sliding liquid becomes more pronounced compared to a single-path channel.

Based on Lemma \ref{Lemma1}, we will analyze the asymptotic performance of MPMP method in the LoS channel.
The details are introduced in Theorem \ref{Theorem1} below.

\begin{theorem}\label{Theorem1}
Under Assumption \ref{Assumption-Mild}, for a LoS channel with an arbitrary CSI delay $\tau$, if the fluid length satisfies $W\lambda > \frac{\lambda}{{\left| {\cos {\theta _{\rm{EOA}}^{{\rm{LoS}}}}} \right|}}$, and the density $\rho$ is large enough, then with three arbitrary samples known, the asymptotic MSE yields:
\begin{equation}
\begin{array}{l}
\mathop {\lim }\limits_{{N_t},{\rho} \to \infty } \left\| {{{{\bf{\hat { h}}}_{m(\tau)}}}(t + \tau) \!-\! {{\bf{h}}_1}(t)} \right\|_2^2 \!= 0.
\end{array}
\!\!\!\!\label{TheoremError}
\end{equation}
\end{theorem}
\begin{proof}
The proof can be found in Appendix B.
\end{proof}
Remarks:
Theorem \ref{Theorem1} indicates that by sliding the FA liquid, the time-varying channel can be transformed to a static channel, when the number of the BS antennas and port density are large enough.
In other words, given a finite number of samples, by sliding the liquid, the FA relieves the Doppler effect caused by UE mobility.
However, Theorem \ref{Theorem1} just analyzes the asymptotic MSE of the LoS path.
In the following, we will study the asymptotic MSE under NLoS channels.

Denote the maximum and minimum values of the NLoS path delay by $\tau_{\rm{max}}$ and $\tau_{\rm{min}}$.
We assume the delay $\tau_p$, the EOA ${\theta _p^{\rm{rx}}}$, the AOA ${\phi _p^{\rm{rx}}}$, and the EOD ${\theta _p}$ are independently and uniformly distributed over the intervals $[\tau_{\rm{min}}, \tau_{\rm{max}}]$, $[0,\pi]$, $[-\pi,\pi]$, and $[0,\pi]$, respectively.
Define four probability density functions (PDFs) of the delay, EOA, AOA, and EOD as $p(\tau _p)$, $p({\theta _p^{\rm{rx}}})$, $p({\phi _p^{\rm{rx}}})$, and $p({\theta _p})$: 
$p({\tau _p}) = {\frac{{{1}}}{{\tau_{\rm{max}}}-\tau_{\rm{min}}}}$, $p({\theta}_p) = p({\theta _p^{\rm{rx}}}) = {\frac{{{1}}}{{\pi}}}$ and $p({\phi _p^{\rm{rx}}}) = {\frac{{{1}}}{{2\pi}}}$.
According to Eq. (\ref{PathGains}), the amplitudes of the NLoS path differ among clusters.
For derivation simplicity, we consider a scenario with a LoS path and a cluster of NLoS paths, where the amplitude of the $p$-th NLoS path is $\beta _p = \frac{1}{{\sqrt {{P_1}} }}$.

Based on Eq. (\ref{3GPPModel}) and Eq. (\ref{ErrorFluidAllAntenna-Derivation}), the MSE is calculated by
\begin{equation}
\begin{array}{l}
\left\| \varepsilon _n \right\|_2^2 = 4(\Xi  + \Omega  + \Lambda),
\end{array}
\!\label{Error-MSE-MultiplePaths}
\end{equation}
where 
\begin{equation}
\begin{array}{l}
\Xi = \frac{{{K_R}}}{{{K_R} + 1}}{\sin ^2}{\varsigma ^{{\rm{LoS}}}} + \frac{1}{{{K_R} + 1}}\frac{1}{{{P_1}}}\sum\limits_{p = 1}^{{P_1}} {} {\sin ^2}{\varsigma _p},
\end{array}
\!\label{Error-MSE-upsilon}
\end{equation}
\begin{equation}
\!\!\!\begin{array}{l}
\!\!\!\!\!\!\Omega =\!\! \frac{{2\sqrt {{K_R}} }}{{{K_R} + 1}}\!\frac{1}{{\sqrt {{P_1}} }}\!\sum\limits_{p = 1}^{{P_1}} {} \sin {\varsigma _p}\sin {\varsigma ^{{\rm{LoS}}}}\!\cos ({\delta _p}\! +\! {\varsigma _p} \!-\! {\delta ^{{\rm{LoS}}}} \!\!-\! {\varsigma ^{{\rm{LoS}}}}),
\end{array}
\!\!\!\!\!\!\!\!\!\label{Error-MSE-xi}
\end{equation}
\begin{equation}
\!\!\!\!\!\!\!\begin{array}{l}
\Lambda  \!=\! \frac{1}{{{K_R} + 1}}\frac{1}{{{P_1}}}\sum\limits_{p = 1}^{{P_1}} {} \!\!\sum\limits_{q = 1,q \ne p}^{{P_1}} {}\!\!\!\! \sin {\varsigma _p}\sin {\varsigma _q}\cos ({\delta _p} \!+\! {\varsigma _p} \!-\! {\delta _q} \!-\! {\varsigma _q}).
\end{array}
\!\!\!\!\!\!\!\!\!\!\!\label{Error-MSE-varpi}
\end{equation}
$\Xi$ includes the non-cross terms of the LoS and NLoS paths, $\Omega$ contains the cross terms of the LoS and NLoS paths, and $\Lambda$ comprises the cross terms of two different NLoS paths.

Then, we will derive the asymptotic performances of $\Xi$, $\Omega $, and $\Lambda$ as follows:
\begin{lemma}\label{Lemma2}
If the number of paths is large enough, the cross term between the LoS and NLoS paths asymptotically yields:
\begin{equation}
\!\begin{array}{l}
\!\!\!\!\!\!\!\! {\cal X} = \mathop {\lim }\limits_{{P_1} \to \infty } \frac{\Omega }{{\sqrt {{P_1}} }} = \frac{{\sqrt {{K_R}} }}{{4\pi f({K_R} + 1)}}\sin {\varsigma ^{{\rm{LoS}}}}\\
\frac{{\cos (2\pi \!{\tau _{\min }} - {\delta ^{{\rm{LoS}}}} - {\varsigma ^{{\rm{LoS}}}}) - \cos (2\pi \!f{\tau _{\max }} \!- {\delta ^{{\rm{LoS}}}} \!- {\varsigma ^{{\rm{LoS}}}})}}{{{\tau _{\max }} - {\tau _{\min }}}}\\
{J_0}(\sqrt {{{D_{n,h}^2}}\! +\! {{D_{n,v}^2}}} \! -\! D_{n,v}){J_0}(\sqrt {{{D_{n,h}^2}} \!+ \!{{D_{n,v}^2}}}  \!+\! D_{n,v})\\
( {J_0}(\frac{{\gamma - (2{C_1} + {F_1})}}{2}) {J_0}(\frac{{\gamma  + (2{C_1} + {F_1})}}{2}) - {J_0}(\frac{{\upsilon  - {F_1}}}{2}){J_0}(\frac{{\upsilon  + {F_1}}}{2})),
\end{array}
\!\!\!\!\!\!\!\!\!\!\!\!\label{Error-crossterm-LoS}
\end{equation}
where ${A_1} = \frac{{\pi \tau {v_x}}}{\lambda }$, ${B_1} = \frac{{\pi \tau {v_y}}}{\lambda }$, ${C_1} = \frac{{(\pi (m - 1)d_v^{{\rm{rx}}} + \pi \tau {v_z})}}{\lambda }$, ${D_1} = \frac{{2\pi t{v_x}}}{\lambda }$, ${E_1} = \frac{{2\pi t{v_y}}}{\lambda }$, ${F_1} = \frac{{2\pi t{v_z}}}{\lambda }$, $D_{n,h} =\frac{\pi {d_h}({n_h} - 1)}{\lambda }$, $D_{n,v} = \frac{\pi {d_v}({n_v} - 1)}{\lambda }$, $\upsilon  =\sqrt {{D_1}^2 + {E_1}^2 + {F_1}^2}$, and $\gamma =  ({{{(2{A_1} + {D_1})}^2} + {{(2{B_1} + {E_1})}^2} + {{(2{C_1} + {F_1})}^2}})^{\frac{1 }{2 }}$ with $v_x$, $v_y$ and $v_z$ being the UE velocity on the $x$, $y$ and $z$ axes.
\end{lemma}
\begin{proof}
The proof can be found in Appendix C.
\end{proof}
Remarks:
Lemma \ref{Lemma2} presents a closed-form expression of the cross term between the LoS and NLoS paths.
In the following, we will derive the closed-form expression of the cross term between two different NLoS paths in Lemma \ref{Lemma3}.
\begin{lemma}\label{Lemma3}
If the number of paths is large enough, the cross term between two different NLoS paths asymptotically yields:
\begin{equation}
\!\!\!\!\!\!\begin{array}{l}
{\cal Y} = \mathop {\lim }\limits_{{P_1} \to \infty } \frac{\Lambda }{{{P_1} - 1}} = \frac{{{{(\sin (2\pi f{\tau _{\max }}) - \sin (2\pi f{\tau _{\min }}))}^2}}}{{16{\pi ^2}{f^2}{{({\tau _{\max }} - {\tau _{\min }})}^2}({K_R} + 1)}}\\
\ \ J_0^2(\!\sqrt {D_{n,h}^2 \!\!+\!\! D_{n,v}^2} \!\!-\!\! {D_{n,v}}) J_0^2(\!\sqrt {D_{n,h}^2 \!\!+\!\! D_{n,v}^2} \!\!+\!\! {D_{n,v}})\\
\ \ {({J_0}(\frac{{\gamma  - (2{C_1} + {F_1})}}{2}){J_0}(\frac{{\gamma  + 2{C_1} + {F_1}}}{2}) \!-\! {J_0}(\frac{{\upsilon  - {F_1}}}{2}){J_0}(\frac{{\upsilon  + {F_1}}}{2}))^2}.
\end{array}
\!\!\!\!\!\!\!\!\!\!\label{CrossTerm-NLoS-clsoed-form}
\end{equation}
\end{lemma}
\begin{proof}
The proof can be found in Appendix D.
\end{proof}
Next, we will derive the closed-form expression of the non-cross terms of LoS and NLoS paths in Lemma \ref{Lemma4}.
\begin{lemma}\label{Lemma4}
If the number of paths is large enough, the non-cross terms of LoS and NLoS paths asymptotically yield:
\begin{equation}
\!\!\!\!\!\begin{array}{l}
{\cal Z} =\mathop {\lim }\limits_{{P_1} \to \infty } \!\!\!\Xi  \! = \frac{{{K_R}}}{{{K_R} + 1}}{\sin ^2}{\varsigma ^{{\rm{LoS}}}} + \frac{1-{J_0}(\eta + {C_1}){J_0}(\eta - {C_1})}{{2({K_R} + 1)}},
\end{array}
\!\!\!\!\!\!\!\!\!\!\label{Term-closed-form}
\end{equation}
where $\eta = \sqrt {{A_1^2} + {B_1^2} + {C_1^2}}$.
\end{lemma}
\begin{proof}
From Eq. (\ref{Error-MSE-upsilon}), we can obtain 
\begin{equation}
\!\!\!\!\!\begin{array}{l}
\mathop {\lim }\limits_{{P_1} \to \infty } \frac{1}{{{P_1}}}\sum\limits_{p = 1}^{{P_1}} {} {\sin ^2}{\varsigma _p} \!=\! E\{ {\sin ^2}{\varsigma _p}\}  \!=\! \frac{1}{2} \!-\! \frac{1}{2}E\{ \cos (2{\varsigma _p})\}.
\end{array}
\!\!\!\!\!\!\!\!\label{Sin2-closed-form}
\end{equation}
Similar to the calculation of $E\left\{ {\cos g_p} \right\}$ in Lemma \ref{Lemma2}, we derive the closed-form expression of $E\left\{ \cos (2{\varsigma _p}) \right\}$ as
\begin{equation}
\!\!\!\!\!\begin{array}{l}
E\left\{ \cos (2{\varsigma _p}) \right\} \!=\! {J_0}(\eta + {C_1}){J_0}(\eta - {C_1}).
\end{array}
\!\!\!\!\!\!\!\!\label{Sin2-closed-form}
\end{equation}
Thus, Lemma \ref{Lemma4} is proved.
\end{proof}
Remarks:
Lemma \ref{Lemma2}, Lemma \ref{Lemma3}, and Lemma \ref{Lemma4} are the basis of Theorem \ref{Theorem2}, where we will derive the lower and upper bounds of the MSE.

\begin{theorem}\label{Theorem2}
Under Assumption \ref{Assumption-Mild}, for the channel with a LoS path and enough NLoS paths, providing that the fluid length satisfies $W\lambda > \frac{\lambda}{{\left| {\cos {\theta _{\rm{EOA}}^{{\rm{LoS}}}}} \right|}}$, three samples are known, and $\rho$ is large enough, the asymptotic MSE yields:
\par \item 1) if $\Lambda  <  - \Omega  + {\cal X}$,
\begin{equation}
\begin{array}{l}
\!\!\!0 \le \!\!\!\mathop {\lim }\limits_{{N_t},{\rho}, P \to \infty } \!\!\!\left\| \varepsilon _n \right\|_2^2 \le \!{\min ( {4({\cal X} +{\cal Y}+{\cal Z} ),{\cal U} } )},
\end{array}
\!\label{TheoremError-MultiplePaths1}
\end{equation}
2) if $\Lambda  >  - \Omega  + {\cal X}$,
\begin{equation}
\begin{array}{l}
\!\!\max (0,4({\cal X} +{\cal Y}+{\cal Z} )) \!\le \!\!\!\mathop {\lim }\limits_{{N_t},{\rho}, P \to \infty } \!\!\!\left\| \varepsilon _n \right\|_2^2 \!\le\! {\cal U},
\end{array}
\!\label{TheoremError-MultiplePaths2}
\end{equation}
3) if $\Lambda  =  - \Omega  + {\cal X}$,
\begin{equation}
\begin{array}{l}
\mathop {\lim }\limits_{{N_t},{\rho}, P \to \infty } \left\| \varepsilon _n \right\|_2^2 = 4({\cal X} +{\cal Y}+{\cal Z} ),
\end{array}
\!\label{TheoremError-MultiplePaths3}
\end{equation}
where
\begin{equation}
\begin{array}{l}
{\cal U} = \frac{2}{{{K_R} + 1}}(1 - {J_0}(\eta  + {C_1}){J_0}(\eta  - {C_1})).
\end{array}
\!\label{TheoremError-UpperMost}
\end{equation}

\end{theorem}
\begin{proof}
The proof can be found in Appendix E.
\end{proof}
Remarks:
For a multi-path channel, Theorem \ref{Theorem2} demonstrates the lower and upper bounds of the prediction MSE, providing that $P \to \infty$.
These MSE bounds are related to the UE velocity and the global optimal port.
In some cases, if the UE velocity and the global optimal port satisfy $\Lambda  <  - \Omega  + {\cal X}$ and $4({\cal X} +{\cal Y}+{\cal Z} ) < {\cal U}$, we may achieve a tighter MSE interval.
For a Ricean channel, if the channel power is mainly concentrated on the LoS path, ${\cal U}$ is a small value, which indicates our proposed method may perform well in the channel with a strong LoS component.

\section{Numerical Results}\label{sec:Simulations}

This section includes the simulation model and numerical results of our proposed scheme.
The carrier frequency is 39 GHz.
We adopt the clustered delay line (CDL) channel model of 3GPP.
The channel model includes 37 paths, which comprises a LoS path and 36 NLoS paths.
The Root Mean Square (RMS) angular spreads of AOD, EOD, AOA, and EOA are $39.01^\circ $, $148.95^\circ $, $-31.93^\circ $, and $31.98^\circ $, respectively.
Table I shows the detailed simulation parameters.
Considering a 3D Urban Macro (3D UMa) scenario, the UEs move at 60 ${\rm{km/h}}$ and 120 ${\rm{km/h}}$.
Each slot contains 14 OFDM symbols, and the duration of a slot is 0.5 $\rm{ms}$.
One time sample is available for each slot.
Each UE sends a sequence of Sounding Reference Signal (SRS) in a time slot.
The antenna configuration is $(\underline M,\underline N)$, where $\underline M$ denotes the number of horizontal antennas, and $\underline N$ is the number of antennas in the vertical direction.
The configurations of the BS antenna are $2 \times 8$, $8 \times 8$, and $32 \times 8$.
The UE has a single FA.
The length of FA is $20 \lambda$ with 300 ports.
The density of ports is 15.
The DL precoder is eigen zero-forcing (EZF) \cite{Sun10TSP}.
We take the DL spectral efficiency (SE) and the DL prediction error as two metrics to assess the prediction method.
\begin{table}[!bht]
\caption{The main simulation parameters.}
\label{table I}
\centering
\begin{tabular}{|p{9em}|p{13em}|}
\hline
Scenario & 3D Urban Macro (3D UMa)\\
\hline
Carrier frequency (GHz) & 39\\
\hline
Number of UEs & 8\\
\hline
BS antenna configuration & $(\underline{M},\underline{N}) = (2,8)$, $(8,8)$, $(32,8)$, $(d_h,d_v)=(0.5\lambda,0.5\lambda)$\\
\hline
UE FA configuration & $(W,M,\rho) = (20,300,15)$\\
\hline
The DL precoder & EZF\\
\hline
Delay spread ($\rm{ns}$) & 616\\
\hline
CSI delay ($\rm{ms}$) & 4\\
\hline
UEs speed (${\rm{km/h}}$) & 60, 120\\
\hline
\end{tabular}
\end{table}

\begin{figure}[!bht]
\centering
\includegraphics[width=2.9in]{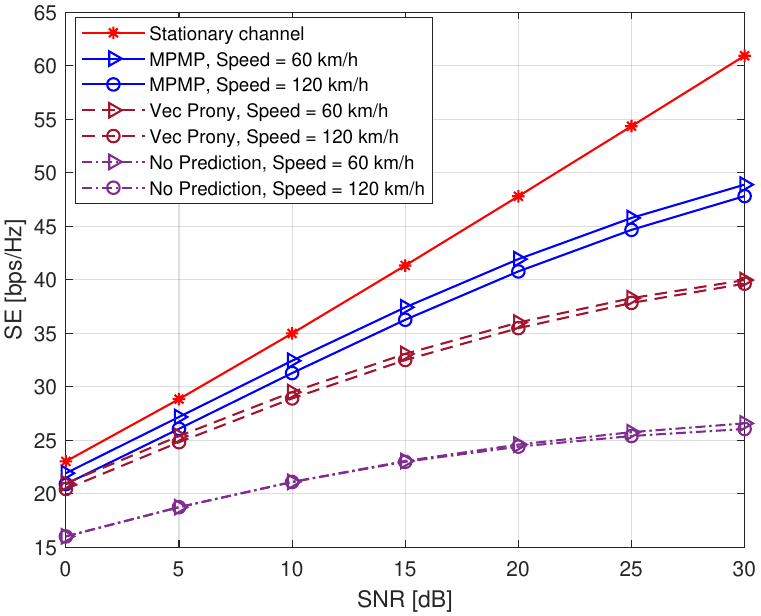}
\caption{The SE versus SNR, the BS has 16 antennas, the CSI delay is 4 $\rm{ms}$, $W=20$, $M=300$.}
\label{fig:SEDifferentSpeed}
\end{figure}
\begin{figure}[!bht]
\centering
\includegraphics[width=2.9in]{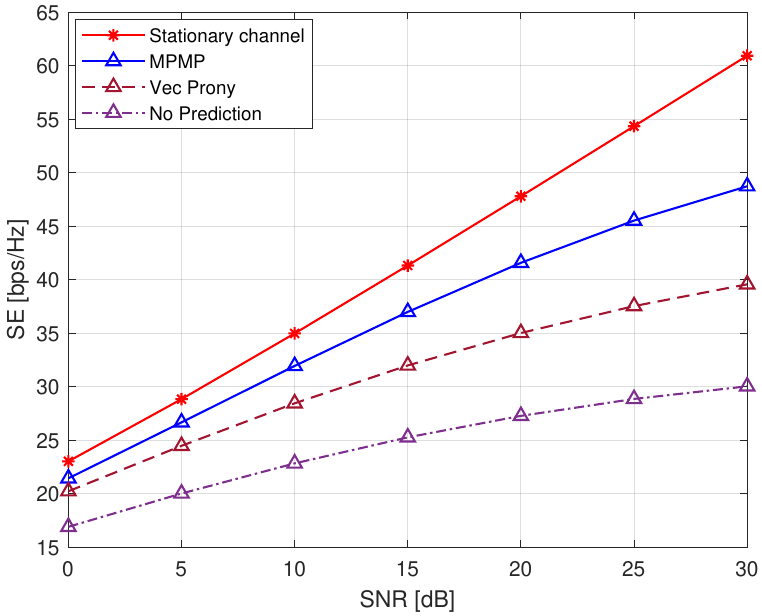}
\caption{The SE versus SNR, the BS has 16 antennas, CSI delay is 4 $\rm{ms}$, $W=20$, $M=300$, multiple velocity levels of UEs, i.e., two at 30 $\rm{km/h}$, two at 60 $\rm{km/h}$, two at 90 $\rm{km/h}$, and two at 120 $\rm{km/h}$.}
\label{fig:SEMultipleSpeeds}
\end{figure}
\begin{figure}[!bht]
\centering
\subfigure[]{
\includegraphics[width=2.9in]{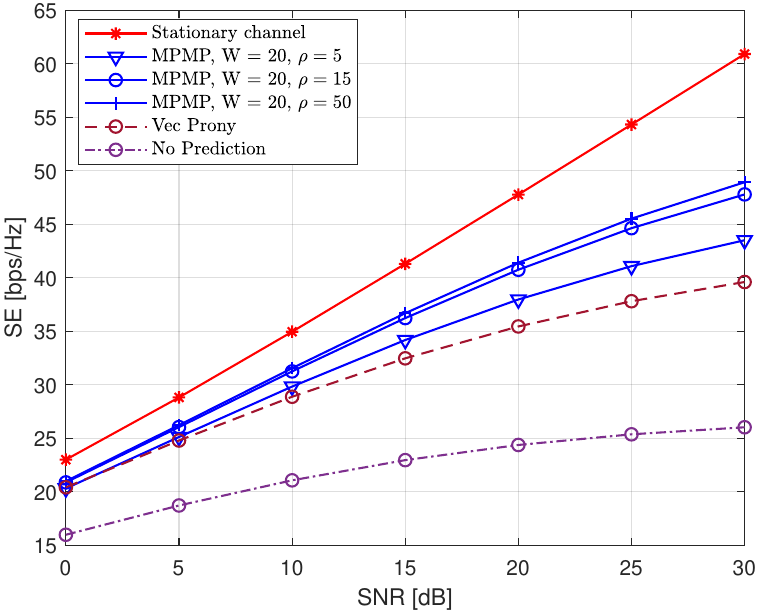}}
\\
\subfigure[]{
\includegraphics[width=2.9in]{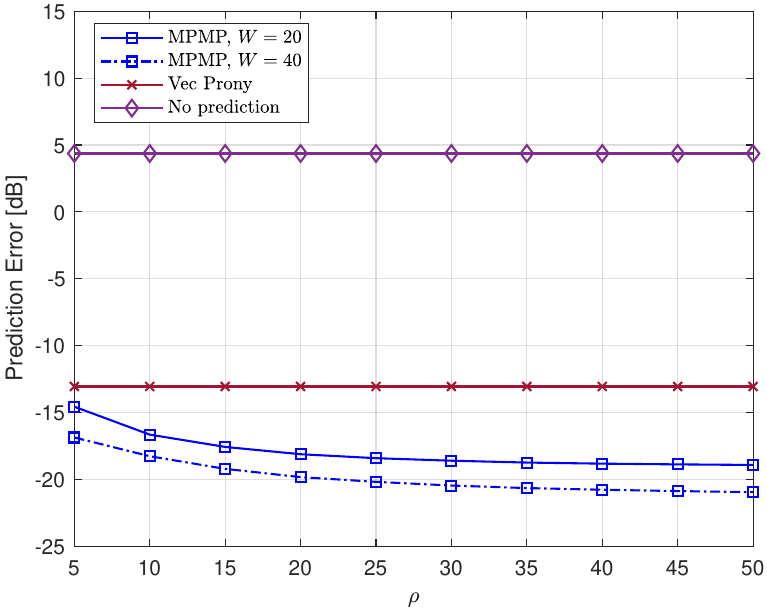}
}
\caption{(a) The SE versus SNR and (b) the prediction error versus $\rho$, the BS has 16 antennas, the CSI delay is 4 $\rm{ms}$, the UEs move at 120 $\rm{km/h}$.}
\label{fig:DifferentPorts}
\end{figure}

Fig. \ref{fig:SEDifferentSpeed} shows the SEs of different methods when the UEs move at 60 ${\rm{km/h}}$ and 120 ${\rm{km/h}}$.
The DL SE is calculated by $\sum\limits_{u = 1}^{{N_{{\rm{UE}}}}} {} E\{ {\rm{log}}_2(1 + {\rm{SINR}}_u)\} $, where ${\rm{SINR}}_u$ is the signal-to-interference-and-noise radio of the $u$-th UE, and ${{N_{{\rm{UE}}}}}$ is the number of UEs.
The expectation is taken over time.
The curve labeled ``Stationary channel" is the ideal setting, which is the upper bound of the performance.
The results without prediction are referred to as ``No prediction".
We select the Vec Prony channel prediction method in \cite{Yin20JSAC} as reference curves.
We may observe that our proposed MPMP method outperforms the performances of the Vec Prony method and the no prediction case in the moderate-mobility and high-mobility scenarios.

Fig. \ref{fig:SEMultipleSpeeds} depicts the SEs of different methods when the UEs move at different speeds, i.e., two at 30 $\rm{km/h}$, two at 60 $\rm{km/h}$, two at 90 $\rm{km/h}$, and two at 120 $\rm{km/h}$.
One may observe that our proposed MPMP method still performs better than the Vec Prony method and the no prediction case.

\begin{figure}[bhtp]
\centering
\subfigure[]{
\includegraphics[width=2.9in]{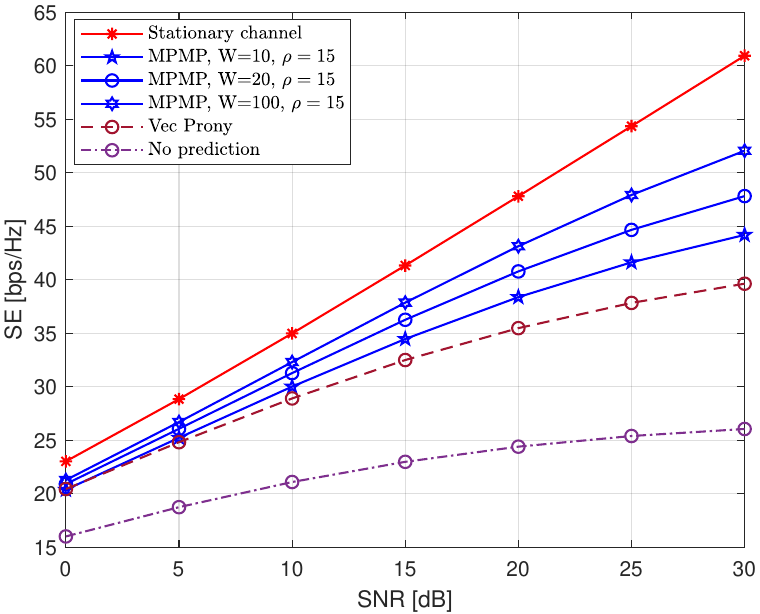}}
\\
\subfigure[]{
\includegraphics[width=2.9in]{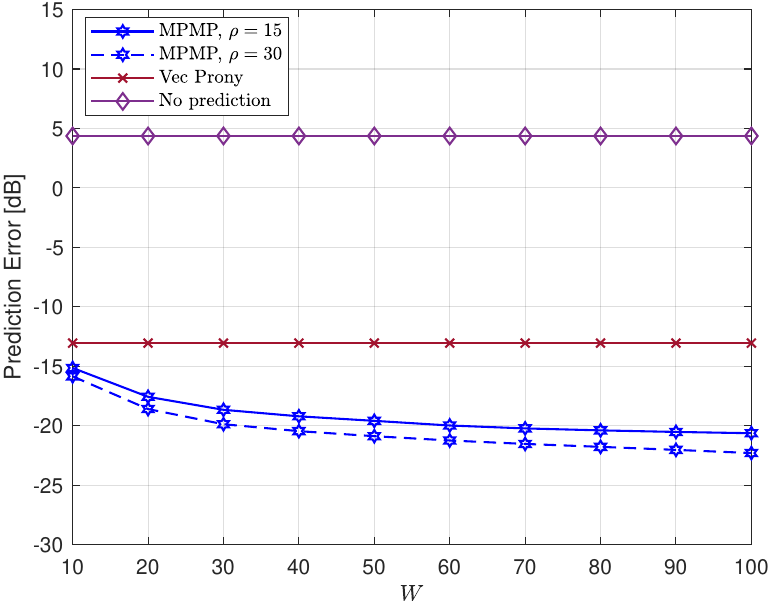}
}
\caption{(a) The SE versus SNR and (b) the prediction error versus $M$, the BS has 16 antennas, the CSI delay is 4 $\rm{ms}$, the UEs move at 120 $\rm{km/h}$.}
\label{fig:DifferentLengths}
\end{figure}

In Fig. \ref{fig:DifferentPorts} (a) and (b), we compare the SEs and prediction errors of different methods when the UE FA has different densities of ports.
The prediction error is defined as $10\log \left\{ {E\left\{ {\frac{{\left\| {{{{\bf{\hat{ h}}}}_{m(\tau)}}(t + \tau ) - {\bf{h}}(t)} \right\|_2^2}}{{\left\| {{\bf{h}}(t)} \right\|_2^2}}} \right\}} \right\}$, where ${{{\bf{\hat{h}}}}_{m(\tau)}}(t+ \tau)$ is the reconstructed channel at time $t+ \tau$ between all BS antennas and the $m(\tau)$-th port of the UE FA.
The expectation is taken over time and UEs.
We may observe that as the density of ports increases, the MPMP method achieves higher SE in Fig. \ref{fig:DifferentPorts} (a), and the prediction accuracy keeps increasing in Fig. \ref{fig:DifferentPorts} (b).

Fig. \ref{fig:DifferentLengths} (a) and (b) depict the SEs and prediction errors of different methods, as the UE FA has different lengths.
It may also be easily observed that with a fixed port density, our MPMP method achieves better performance, as the liquid length increases.
The reason is that a longer liquid covers a larger potential range, provided ${{W\lambda }} < L_{{\varepsilon}}$.
From Fig. \ref{fig:DifferentLengths} (b), it is easily observed that the prediction error of the MPMP method decreases as the FA length increases from $10\lambda$ to $100\lambda$.
\begin{figure}[bhtp]
\centering
\includegraphics[width=2.9in]{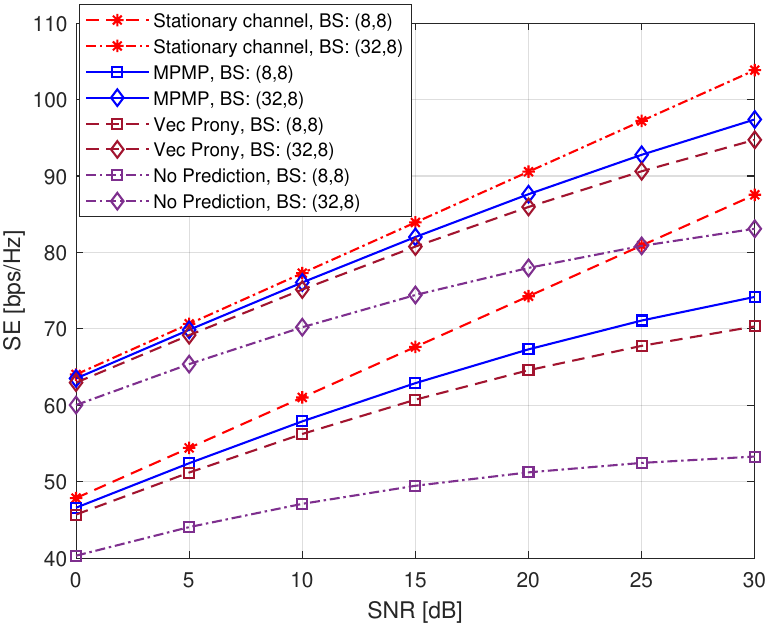}
\caption{The SE versus SNR, CSI delay is 4 $\rm{ms}$, the UEs move at 120 $\rm{km/h}$, $W=20$, $M=300$.}
\label{fig:DifferentAntennas}
\end{figure}

In Fig. \ref{fig:DifferentAntennas}, we compare the SEs of different methods when the BS antenna configurations are $(8,8)$ and $(32,8)$, respectively.
One may observe that our proposed MPMP method still performs well when the BS has larger number of antennas.

\section{Conclusion}\label{sec:Conclusion}

In this paper, we addressed the mobility problem with the assistance of FA. We proposed a matrix pencil-based moving port prediction method, which transforms the time-varying channel to the static channel by sliding the liquid. We proposed a port selection method to obtain the optimal port. In the theoretical analysis, we derived the minimum FA length and the maximum sliding speed. For a LoS channel, we also proved that the prediction MSE converges to zero if the density of ports and the number of BS antennas are large enough. For a multi-path channel, we also derived the closed-form lower and upper bounds of the prediction MSE. Simulation results demonstrate that, with the assistance of FA, our proposed MPMP method provides significant gain in the high-mobility scenario.

\appendix

\subsection{Proof of Lemma \ref{Lemma1}}\label{Lemma: speed of sluiding fluid}
If $W\lambda  \ge \frac{{\lambda }}{{\left| {\cos \theta _{{\rm{EOA}}}^{{\rm{LoS}}}} \right|}}$, based on Eq. (\ref{Port-LoSk}), Eq. (\ref{Port-LoSkA}) and Eq. (\ref{Port-LoSkB}), we can obtain ${\bar A}<{\bar B}-1$.
Therefore, in Eq. (\ref{Port-LoS}), the range of $k(\Delta t)$ includes at least one integer.

Denote two neighboring samples at time $t_0$ and $t_0+T$ by $h_{m}^{{\rm{LoS}}}(t_0)$ and $h_{m}^{{\rm{LoS}}}(t_0+T)$.
The corresponding optimal ports are ${m}(t_0)$ and ${m}(t_0+T)$, respectively.
According to Eq. (\ref{Port-LoS}), the sliding speed $v_f$ is calculated by
\begin{equation}
\begin{array}{l}
{v_f} \!=\! \frac{{d_v^{\rm{rx}}}{\left| {{m}( {t_0+T}) - {m}( {t_0})} \right|}}{{T }} \!\approx\! \left| {\left[ {\frac{{\lambda \omega ^{{\rm{LoS}}}}}{{\cos \theta _{\rm{EOA}}^{{\rm{LoS}}}}} + \frac{{\lambda ({k({{t_0} + T})} - {k({{t_0}})})}}{T{\cos \theta _{\rm{EOA}}^{{\rm{LoS}}}}}} \right]} \right|
\end{array}.
\label{Speed-PortLoS-delta-t}
\end{equation}
Let ${k({{t_0}})} \le {k({{t_0} + T})}$.
In such case, ${k({{t_0} + T})} - {k({{t_0}})} = 0$ or ${k({{t_0} + T})} - {k({{t_0}})} = 1$.
Eventually, when ${k({{t_0} + T})} - {k({{t_0}})} = 0$, the speed is
\begin{equation}
\begin{array}{l}
{v_f} \approx \left| {\left[ {\frac{{\lambda \omega ^{{\rm{LoS}}}}}{{\cos \theta _{\rm{EOA}}^{{\rm{LoS}}}}} } \right]} \right|.
\end{array}
\!\label{Speed-PortLoS-delta-t}
\end{equation}
As ${k({{t_0} + T})} - {k({{t_0}})} = 1$, the speed is
\begin{equation}
\begin{array}{l}
{v_f} \approx \left| {\left[ {\frac{{\lambda \omega ^{{\rm{LoS}}}}T+ {\lambda}}{T{\cos \theta _{\rm{EOA}}^{{\rm{LoS}}}}}} \right]} \right|.
\end{array}
\!\label{Speed-PortLoS-delta-t}
\end{equation}
Due to $\max({{m}(t_0+T), {m}(t_0)}) \le M$, we can obtain ${v_f} \le \frac{{W\lambda }}{T}$.
Let ${v_f}(t_1) = \min{(\left| {\left[ {\frac{{\lambda \omega ^{{\rm{LoS}}}}}{{\cos \theta _{\rm{EOA}}^{{\rm{LoS}}}}} } \right]} \right|, \left| {\left[ {\frac{{\lambda \omega ^{{\rm{LoS}}}}T+ {\lambda}}{T{\cos \theta _{\rm{EOA}}^{{\rm{LoS}}}}}} \right]} \right|)}$ and ${v_f}(t_2) = \max{(\left| {\left[ {\frac{{\lambda \omega ^{{\rm{LoS}}}}}{{\cos \theta _{\rm{EOA}}^{{\rm{LoS}}}}} } \right]} \right|, \left| {\left[ {\frac{{\lambda \omega ^{{\rm{LoS}}}}T+ {\lambda}}{T{\cos \theta _{\rm{EOA}}^{{\rm{LoS}}}}}} \right]} \right|)}$.

Thus, the Lemma \ref{Lemma1} is proved.

\subsection{Proof of Theorem \ref{Theorem1}}\label{appendix:Error of the single LoS}

Due to $W\lambda > \frac{\lambda}{{\left| {\cos {\theta _{\rm{EOA}}^{{\rm{LoS}}}}} \right|}}$, the port is selected as Eq. (\ref{Port-LoS}).
Denote the optimal port at time $t+\tau$ by $m(\tau)$.
Then, we slide the liquid to the $m(\tau)$-th port, and get the channel between the $n$-th BS antenna and the UE FA as
\begin{equation}
\begin{array}{l}
{ h_{n,m(\tau)}}(t + \tau) = \sqrt {\frac{{{K_R}}}{{{K_R} + 1}}} {e^{j2\pi f{\tau ^{\rm{LoS}}}}} {e^{\frac{{j2\pi {{({{\bf{r}}^{\rm{tx},{\rm{LoS}}}})}^T}{{\bf{d}}_n^{\rm{tx}}}}}{\lambda }}}\\
\ \ \ \ \ \ \ \ \ \ \ \ \ \ {e^{j2\pi {\omega ^{{\rm{LoS}}}}(t + \tau)}} {e^{j\frac{{2\pi }}{\lambda }{d_v^{\rm{rx}}}\cos \theta _{\rm{EOA}}^{{\rm{LoS}}}(m(\tau) - 1)}}.
\end{array}
\!\label{FluidChannelMn-deltat}
\end{equation}
The error between ${ h_{n,m(\tau)}}(t + \tau)$ and ${ h_{n,1}}(t)$ is calculated by
\begin{equation}
\!\!\!\!\!\!\!\!\!\!\!\!\!\!\!\!\begin{array}{l}
\!\!{\varepsilon _n} \!=\! {h_{n,m}}(t + \tau ) \!-\! {h_{n,1}}(t) \!=\!\! \sqrt {\frac{K_R}{{K_R + 1}}} {e^{j2\pi f{\tau ^{{\rm{LoS}}}}}}\!\!{e^{\frac{{j2\pi {{({{\bf{r}}^{{\rm{tx,LoS}}}})}^T}{\bf{d}}_n^{{\rm{tx}}}}}{\lambda }}}\\
{e^{j2\pi {\omega ^{{\rm{LoS}}}}t}}({e^{j2\pi {\omega ^{{\rm{LoS}}}}\tau }}{e^{j\frac{{2\pi }}{\lambda }d_v^{{\rm{rx}}}\cos \theta _{{\rm{EOA}}}^{{\rm{LoS}}}(m(\tau ) - 1)}} - 1).
\end{array}
\!\!\!\!\!\!\!\!\!\!\!\!\!\!\!\!\!\!\!\!\!\!\label{Error-n}
\end{equation}
According to Eq. (\ref{Error-n-ALoS}) and Eq. (\ref{Port-LoS}), we calculate the MSE $\left\| {{\varepsilon _n}} \right\|_2^2$ by
\begin{equation}
\begin{array}{l}
\left\| {{\varepsilon _n}} \right\|_2^2 \!=\!\! {\frac{{{K_R}}}{{{K_R} + 1}}}\left\| {{e^{g(y)}} \!-\! 1} \right\|_2^2 \!=\!\!{\frac{{{K_R}}}{{{K_R} + 1}}} (2 \!-\! 2\cos \left( g(y) \right)),
\end{array}
\!\!\!\label{Error-n}
\end{equation}
where
\begin{equation}
\begin{array}{l}
g(y) = \frac{{2\pi}}{\rho}\cos \theta _{\rm{EOA}}^{{\rm{LoS}}}y,
\end{array}
\!\label{gy}
\end{equation}
and
\begin{equation}
\begin{array}{l}
y = \lambda \frac{{\bmod (\omega ^{{\rm{LoS}}}\tau,1 )}}{{ {d_v^{\rm{rx}}}\cos \theta _{\rm{EOA}}^{{\rm{LoS}}}}} - \left[ {\lambda \frac{{\bmod ( \omega ^{{\rm{LoS}}}\tau,1 )}}{{{d_v^{\rm{rx}}}\cos \theta _{\rm{EOA}}^{{\rm{LoS}}}}}} \right].
\end{array}
\!\label{y}
\end{equation}
The range of $y$ is
\begin{equation}
\begin{array}{l}
y \in ( - 0.5,0.5]
\end{array}.
\!\label{y-range}
\end{equation}
Due to $\theta _{\rm{EOA}}^{{\rm{LoS}}} \in [0,\frac{\pi }{2}) \cup (\frac{\pi }{2},\pi ]$, the range of $g(y)$ is $[-\frac{{\pi}}{\rho},0) \cup (0,\frac{{\pi}}{\rho}]$.
As $\rho \to \infty$, the asymptotic upper bound of ${\left\| {{\varepsilon _n}} \right\|_2^2}$ is calculated by
\begin{equation}
\begin{array}{l}
\mathop {\lim }\limits_{\rho \to \infty } {\left\| {{\varepsilon _n}} \right\|_2^2}_{\max} = \mathop {\lim }\limits_{\rho \to \infty } \frac{{{K_R}}}{{{K_R} + 1}}(2 - 2\cos(\frac{{\pi}}{\rho})) = 0
\end{array}.
\!\label{Error-upper-bound}
\end{equation}
Since ${\left\| {{\varepsilon _n}} \right\|_2^2} \ge 0$, we obtain ${\left\| {{\varepsilon _n}} \right\|_2^2} = 0$.
Based on Eq. (\ref{AllErrorFluidAllAntenna}), we obtain
\begin{equation}
\begin{array}{l}
\mathop {\lim }\limits_{{N_t},\rho \to \infty }\!\!\! {\left\| {\bm{\varepsilon }} \right\|_2^2} \!=\!\! \mathop {\lim }\limits_{{N_t},\rho \to \infty }\! \left\| {{{{\bf{\hat {h}}}_{m(\tau)}}}(t + \tau ) - {{\bf{h}}_1}(t)} \right\|_2^2\\
\ \ \ \ \ \ \ \ \ \ \ \ \ \ = \mathop {\lim }\limits_{{N_t},\rho \to \infty } \sum\limits_{n = 1}^{{N_t}} {} {\left\| {{\varepsilon _n}} \right\|_2^2} =0.
\end{array}
\!\label{TheoremError-nall}
\end{equation}
Thus, Theorem \ref{Theorem1} is proved.

\subsection{Proof of Lemma \ref{Lemma2}}\label{Lemma3: error of corss term of LoS and NLoS}

Based on Eq. (\ref{Error-MSE-xi}), $\frac{\Omega }{{\sqrt P_1 }}$ is calculated by
\begin{equation}
\begin{array}{l}
\!\!\!\!\!\!\!\!\!\!\!\!\mathop {\lim }\limits_{P_1 \to \infty }\!\!\!\frac{\Omega }{{\sqrt P_1 }} \!\!= \!\! \frac{{\sqrt {{K_R}} }}{{{K_R} + 1}}\!\!\mathop {\lim }\limits_{{P_1} \to \infty } \!\sum\limits_{p = 1}^{{P_1}} {}\!\! \frac{\sin \!{\varsigma _p}\!\sin \!{\varsigma ^{{\rm{LoS}}}}\!\!\cos ({\delta _p} \!+\! {\varsigma _p} \!-\! {\delta ^{{\rm{LoS}}}} \!-\! {\varsigma ^{{\rm{LoS}}}})}{P_1}\\
\!\!\!\!\!\!\!\!\approx \frac{{\sqrt {{K_R}} }}{{{K_R} + 1}}\sin {\varsigma ^{{\rm{LoS}}}}E\{ \sin {\varsigma _p}\cos ({\delta _p} + {\varsigma _p} - {\delta ^{{\rm{LoS}}}} - {\varsigma ^{{\rm{LoS}}}})\},
\end{array}
\!\!\!\!\!\!\!\!\!\!\!\!\!\!\!\!\label{Error-Xi-P}
\end{equation}
where the expectation operation is averaged over multiple paths.
Since the EOD, EOA, AOA, and delay are distributed independently, we have
\begin{equation}
\!\!\!\!\!\!\begin{array}{l}
\!\!\!\!\!\!\!\!\!\!E\{ \sin {\varsigma _p}\cos ({\delta _p} + {\varsigma _p} - {\delta ^{{\rm{LoS}}}} - {\varsigma ^{{\rm{LoS}}}})\} \\
=\frac{1}{2}E\{ \sin k_p\} ((E\{ \cos a_p\}  - E\{ \cos g_p\} )E\{ \cos b_p\} \\
 + (E\{ \sin g_p\}  - E\{ \sin a_p\} )E\{ \sin b_p\} )\\
 + \frac{1}{2}E\{ \cos k_p \} ((E\{ \sin a_p\}  - E\{ \sin g_p\} )E\{ \cos b_p\} \\
 + (E\{ \cos a_p\}  - E\{ \cos g_p\} )E\{ \sin b_p\} ),
\end{array}
\!\!\!\!\!\!\!\!\!\!\!\!\label{Error-Xi-P-Expectation}
\end{equation}
with ${k_p} = 2\pi f{\tau _p} - {\delta ^{{\rm{LoS}}}} - {\varsigma ^{{\rm{LoS}}}}$, ${b_p} = 2{D_{n,h}}\sin {\theta _p}\sin {\phi _p} + 2{D_{n,v}}\cos {\theta _p}$, and
\begin{equation}
\!\!\!\!\!\!\!\!\!\!\begin{array}{l}
{a_p} \!=\! (2{A_1} \!+\! {D_1})\!\sin\theta _p^{{\rm{rx}}}\!\cos\phi _p^{{\rm{rx}}} \!+\!\! (2{B_1} \!+\!\! {E_1})\!\sin\theta _p^{{\rm{rx}}}\!\sin \phi _p^{{\rm{rx}}} \\
\ \ \ \ + (2{C_1} \!+ \!{F_1})\cos\theta _p^{{\rm{rx}}},
\end{array}
\!\!\!\!\!\!\!\!\!\!\!\!\!\!\!\!\label{Error-Xi-P-Parameter-a}
\end{equation}
\begin{equation}
\!\!\!\!\!\!\!\!\!\!\begin{array}{l}
{g_p} \!=\! {D_1}\sin\theta _p^{{\rm{rx}}}\cos\phi _p^{{\rm{rx}}}\! +\! {E_1}\sin \theta _p^{{\rm{rx}}}\sin \phi _p^{{\rm{rx}}} \!+ \!{F_1}\cos \theta _p^{{\rm{rx}}}.
\end{array}
\!\!\!\!\!\!\!\!\!\!\!\!\!\!\!\!\label{Error-Xi-P-Parameter-g}
\end{equation}
The derivation of Eq. (\ref{Error-Xi-P}) is transformed into the calculations of $E\left\{ {\cos a_p} \right\}$, $E\left\{ {\sin a_p} \right\}$, $E\left\{ {\cos b_p} \right\}$, $E\left\{ {\sin b_p} \right\}$, $E\left\{ {\cos k_p} \right\}$, $E\left\{ {\sin k_p} \right\}$, $E\left\{ {\cos g_p} \right\}$, and $E\left\{ {\sin g_p} \right\}$.
In the following, we will take the calculation of $E\left\{ {\cos g_p} \right\}$ as an example:
\begin{equation}
\!\!\!\!\!\begin{array}{l}
E\{ \cos {g_p}\}  = \int_0^\pi  {} \frac{1}{\pi }\cos({F_1}\cos\theta _p^{{\rm{rx}}}){\rm{d}}\theta _p^{{\rm{rx}}}\\
\ \ \ \ \int_0^{2\pi } {} \!\!\frac{1}{{2\pi }}\!\!\cos(\!{D_1}\!\sin\theta _p^{{\rm{rx}}}\!\cos\phi _p^{{\rm{rx}}} \!+\! {E_1}\!\sin\theta _p^{{\rm{rx}}}\!\sin \phi _p^{{\rm{rx}}}){\rm{d}}\phi _p^{{\rm{rx}}} \\
\ \ \ \ \ - \int_0^\pi  {} \frac{1}{\pi }\sin ({F_1}\cos\theta _p^{{\rm{rx}}}){\rm{d}}\theta _p^{{\rm{rx}}}\\
\ \ \ \int_0^{2\pi } {} \frac{1}{{2\pi }}\sin ({D_1}\sin\theta _p^{{\rm{rx}}}\cos\phi _p^{{\rm{rx}}} \!\!+\!\! {E_1}\sin\theta _p^{{\rm{rx}}}\sin \phi _p^{{\rm{rx}}}){\rm{d}}\phi _p^{{\rm{rx}}},
\end{array}
\!\!\!\!\label{Error-Xi-P-E-COS}
\end{equation}
By using the integration formulas in \cite{TableIntegral}
\begin{equation}
\begin{array}{l}
\!\!\!\!\int_0^{2\pi } {} {e^{{\bar p}\cos {\bar x} + {\bar q}\sin {\bar x}}}\sin ({\bar a}\cos {\bar x} + {\bar b}\sin {\bar x} - {\bar m}{\bar x}){\rm{d}}{\bar x}\\
\!\!\!\!= i\pi {[{({\bar b} -\! {\bar p})^2} + \!{({\bar a} + \!{\bar q})^2}]^{ - \frac{{\bar m}}{2}}}\{ {(\underline{A} + i\underline{B})^{\frac{m}{2}}}{I_m}(\sqrt {\underline{C} \!-\! i\underline{D}} ) \\
 - {(\underline{A} - i\underline{B})^{\frac{{\bar m}}{2}}}{I_{\bar m}}(\sqrt {\underline{C} + i\underline{D}} )\},
\end{array}
\!\!\!\label{Integration-formula-3}
\end{equation}
and
\begin{equation}
\begin{array}{l}
\!\!\!\!\int_0^{2\pi } {} {e^{{\bar p}\cos {\bar x} + {\bar q}\sin {\bar x}}}\cos ({\bar a}\cos {\bar x} + {\bar b}\sin {\bar x} - {\bar m}{\bar x}){\rm{d}}{\bar x}\\
\!\!\!\! = i\pi {[{({\bar b} -\! {\bar p})^2} +\! {({\bar a} +\! {\bar q})^2}]^{ - \frac{{\bar m}}{2}}}\{ {(\underline{A} + i\underline{B})^{\frac{m}{2}}}{I_m}(\sqrt {\underline{C} \!- \!i\underline{D}} ) \\
 + {(\underline{A} - i\underline{B})^{\frac{{\bar m}}{2}}}{I_{\bar m}}(\sqrt {\underline{C} + i\underline{D}} )\},
\end{array}
\!\!\!\label{Integration-formula-4}
\end{equation}
where ${({\bar b} - {\bar p})^2} + {({\bar a} + {\bar q})^2} > 0$, $\underline{A} = {{\bar p}^2} - {{\bar q}^2} + {{\bar a}^2} - {{\bar b}^2}$, $\underline{B} = 2({\bar p}{\bar q} + {\bar a}{\bar b})$, $\underline{C} = {{\bar p}^2} + {{\bar q}^2} - {{\bar a}^2} - {{\bar b}^2}$, and $\underline{D} = 2({\bar a}{\bar p} + {\bar b}{\bar q})$, and setting $\underline{A} = (D_1^2 - E_1^2)\sin^2\theta _p^{{\rm{rx}}}$, $\underline{B} = 2{D_1}{E_1}\sin^2\theta _p^{{\rm{rx}}}$, $\underline{C} =  - (D_1^2 + E_1^2)\sin^2\theta _p^{{\rm{rx}}}$ and $\underline{D}=0$, we can write
\begin{equation}
\!\!\!\!\!\!\begin{array}{l}
\int_0^{2\pi } {} \cos({D_1}\sin\theta _p^{{\rm{rx}}}\cos\phi _p^{{\rm{rx}}} + {E_1}\sin\theta _p^{{\rm{rx}}}\sin \phi _p^{{\rm{rx}}}){\rm{d}}\phi _p^{{\rm{rx}}}\\
 = 2\pi {I_0}(i\sqrt {D_1^2 + E_1^2} \sin\theta _p^{{\rm{rx}}})
 \mathop  = \limits^{(c)} {J_0}({\sqrt {{D_1^2 + E_1^2}}} \sin {\theta _p^{\rm{rx}}} ),
\end{array}
\!\!\!\!\!\label{Error-Xi-P-E-COS-Result1}
\end{equation}
\begin{equation}
\!\!\!\!\!\begin{array}{l}
\int_0^{2\pi } {}\!\! \sin({D_1}\sin\theta _p^{{\rm{rx}}}\cos\phi _p^{{\rm{rx}}}\! +\! {E_1}\sin\theta _p^{{\rm{rx}}}\sin \phi _p^{{\rm{rx}}}){\rm{d}}\phi _p^{{\rm{rx}}}\!=\!0,
\end{array}
\!\!\!\!\label{Error-Xi-P-E-COS-Result2}
\end{equation}
where $(c)$ is derived by ${I_n}(z) = {i^{ - n}}{J_n}(iz)$ and ${J_0}(z) = {J_0}(-z)$.
Then, 
\begin{equation}
\!\!\!\!\!\!\!\begin{array}{l}
E\{ \cos {g_p}\}  \!\!= \!\!\int_0^\pi  {}\!\! \frac{1}{\pi }\!\cos(\!{F_1}\!\cos\theta _p^{{\rm{rx}}}){J_0}(\!\sqrt {D_1^2 \!+\! E_1^2} \sin\theta _p^{{\rm{rx}}}){\rm{d}}\theta _p^{{\rm{rx}}}\\
\ \ \ \ \ \ \ \ \ \ \ =\!\!\int_0^{\frac{\pi }{2}}  {}\!\! \frac{2}{\pi }\!\cos(\!{F_1}\!\cos\theta _p^{{\rm{rx}}}){J_0}(\!\sqrt {D_1^2 \!+\! E_1^2} \sin\theta _p^{{\rm{rx}}}){\rm{d}}\theta _p^{{\rm{rx}}}.
\end{array}
\!\!\!\!\!\!\label{Error-Xi-P-E-COS-derive}
\end{equation}
According to the integral formula in \cite{TableIntegral}
\begin{equation}
\begin{array}{l}
\!\!\!\int_0^{\frac{\pi }{2}}  {J_{\bar \nu }}({\bar \mu} {\bar z}\sin {\bar t})\cos({\bar \mu} {\bar x}\cos {\bar t}){\rm{d}}{\bar t} \\
\ \ \ \ \ = \frac{\pi }{2}{J_{\frac{{\bar \nu } }{2}}}({\bar \mu} \frac{{\sqrt {{{\bar x}^2} + {{\bar z}^2}} + {\bar x}}}{2}){J_{\frac{{\bar \nu } }{2}}}({\bar \mu} \frac{{\sqrt {{{\bar x}^2} + {{\bar z}^2}} - {\bar x}}}{2}),
\end{array}
\!\label{Integration-formula-6}
\end{equation}
and letting ${\bar \mu}=1$, ${\bar z} = \sqrt {D_1^2 + E_1^2}$, ${\bar x}={F_1}$, and ${\bar \nu}=0$, we can write
\begin{equation}
\!\!\!\!\!\!\!\begin{array}{l}
E\{ \cos {g_p}\} ={J_0}(\frac{{\upsilon  + {F_1}}}{2}){J_0}(\frac{{\upsilon  - {F_1}}}{2}).
\end{array}
\!\!\!\!\!\!\label{Error-Xi-P-E-COS-ResultFinal}
\end{equation}
Likewise, we derive $E\left\{ {\cos a_p} \right\}$, $E\left\{ {\sin a_p} \right\}$, $E\left\{ {\cos b_p} \right\}$, $E\left\{ {\sin b_p} \right\}$, $E\left\{ {\cos k_p} \right\}$, $E\left\{ {\sin k_p} \right\}$, and $E\left\{ {\sin g_p} \right\}$ as follows:
\begin{equation}
\!\!\!\!\!\!\!\begin{array}{l}
E\{ \cos{a_p}\} \!= {J_0}(\frac{{\gamma  + 2{C_1} + {F_1}}}{2}) {J_0}(\frac{{\gamma  - (2{C_1} + {F_1})}}{2}),
\end{array}
\!\!\!\!\!\!\label{Error-Xi-P-E-COSa-Result}
\end{equation}
\begin{equation}
\!\!\!\!\!\!\!\begin{array}{l}
E\{ \!\cos {b_p}\!\}\!\! =\!\! {J_0}(\!\sqrt {D_{n,h}^2 \!\!+\!\! D_{n,v}^2}  \!\!+\!\! {D_{n,v}}){J_0}(\!\sqrt {D_{n,h}^2 \!+ \!D_{n,v}^2}\! -\! {D_{n,v}}),
\end{array}
\!\!\!\!\!\!\label{Error-Xi-P-E-COSb-Result}
\end{equation}
\begin{equation}
\!\!\!\!\!\!\!\begin{array}{l}
E\{ \sin{a_p}\} = E\{ \sin{b_p}\}= E\{ \sin{g_p}\} =0,
\end{array}
\!\!\!\!\!\!\label{Error-Xi-P-E-SINa-Result}
\end{equation}
\begin{equation}
\!\!\!\!\!\!\!\begin{array}{l}
E\{ \cos{k_p}\} \!\!=\!\! \int_{{\tau _{\min }}}^{{\tau _{\max }}} {} \!\frac{\cos (2\pi f{\tau _p} - {\delta ^{{\rm{LoS}}}} - {\varsigma ^{{\rm{LoS}}}})}{{{\tau _{\max }} - {\tau _{\min }}}}{\rm{d}}{\tau _p}\\
\ \ \ \ \ \ \ \ = \frac{{\sin (2\pi f{\tau _{\max }} - {\delta ^{{\rm{LoS}}}} \!-\! {\varsigma ^{{\rm{LoS}}}}) - \sin (2\pi f{\tau _{\min }} - {\delta ^{{\rm{LoS}}}} \!-\! {\varsigma ^{{\rm{LoS}}}})}}{{2\pi f({\tau _{\max }} \!- {\tau _{\min }})}},
\end{array}
\!\!\!\!\!\!\label{Error-Xi-P-E-COSk-Result}
\end{equation}
\begin{equation}
\!\!\!\!\!\!\!\begin{array}{l}
E\{ \sin {k_p}\}  = \int_{{\tau _{\min }}}^{{\tau _{\max }}} {} \frac{{\sin (2\pi f{\tau _p} - {\delta ^{{\rm{LoS}}}} - {\varsigma ^{{\rm{LoS}}}})}}{{{\tau _{\max }} - {\tau _{\min }}}}{\rm{d}}{\tau _p}\\
\ \ \ \ \ \ \ \ = \frac{{\cos (2\pi f{\tau _{\min }} - {\delta ^{{\rm{LoS}}}} \!- \!{\varsigma ^{{\rm{LoS}}}}) - \cos (2\pi f{\tau _{\max }} - {\delta ^{{\rm{LoS}}}} \!-\! {\varsigma ^{{\rm{LoS}}}})}}{{2\pi f({\tau _{\max }} \!- {\tau _{\min }})}}.
\end{array}
\!\!\!\!\!\!\label{Error-Xi-P-E-SINk-Result}
\end{equation}
Finally, we derive the closed-form expression of $\mathop {\lim }\limits_{P_1 \to \infty } \frac{\Omega }{{\sqrt P_1 }}$ as:

\begin{equation}
\begin{array}{l}
\!\!\!\!\!\!\mathop {\lim }\limits_{P_1 \to \infty }\!\!\!\frac{\Omega }{{\sqrt P_1 }} \!\!= \frac{{\sqrt {{K_R}} }}{{4\pi f({K_R} + 1)}}\sin {\varsigma ^{{\rm{LoS}}}}\\
\frac{{\cos (2\pi f{\tau _{\min }} - {\delta ^{{\rm{LoS}}}} - {\varsigma ^{{\rm{LoS}}}}) - \cos (2\pi f{\tau _{\max }} - {\delta ^{{\rm{LoS}}}} - {\varsigma ^{{\rm{LoS}}}})}}{{({\tau _{\max }} - {\tau _{\min }})}}\\
{J_0}(\sqrt {D_{n,h}^2 + D_{n,v}^2}  + {D_{n,v}}){J_0}(\sqrt {D_{n,h}^2 + D_{n,v}^2}  - {D_{n,v}})\\
({J_0}(\frac{{\gamma  + 2{C_1} + {F_1}}}{2}){J_0}(\frac{{\gamma  - (2{C_1} + {F_1})}}{2}) - {J_0}(\frac{{\upsilon  + {F_1}}}{2}){J_0}(\frac{{\upsilon  - {F_1}}}{2})).
\end{array}
\!\!\!\!\!\!\!\!\!\!\!\!\!\!\!\!\label{Error-Xi-P-DerivedResult}
\end{equation}
Thus, Lemma \ref{Lemma2} is proved.

\subsection{Proof of Lemma \ref{Lemma3}}\label{Lemma4: error of corss term of two different NLoS paths}

According to Eq. (\ref{Error-MSE-varpi}), we compute $\mathop {\lim }\limits_{{P_1} \to \infty } \frac{\Lambda }{{{P_1} - 1}}$ as:
\begin{equation}
\!\!\!\!\!\!\!\begin{array}{l}
\mathop {\lim }\limits_{{P_1} \to \infty } \frac{\Lambda }{{{P_1} - 1}} = \mathop {\lim }\limits_{{P_1} \to \infty } \frac{1}{{{K_R} + 1}}\frac{1}{{P_1({P_1} - 1)}}\\
\ \ \ \ \ \ \sum\limits_{p = 1}^{{P_1}} {} \sum\limits_{q = 1,q \ne p}^{{P_1}} {} \sin {\varsigma _p}\sin {\varsigma _q}\cos ({\delta _p} + {\varsigma _p} - {\delta _q} - {\varsigma _q})\\
\ \ \ \ \ \ \approx \frac{1}{{{K_R} + 1}}E\left\{ {\sin {\varsigma _p}\sin {\varsigma _q}\cos ({\delta _p} + {\varsigma _p} - {\delta _q} - {\varsigma _q})} \right\},
\end{array}
\!\!\!\!\!\!\!\!\!\!\!\!\label{Error-Omega-P}
\end{equation}
where
\begin{equation}
\!\!\!\!\!\!\begin{array}{l}
\!\!\!\!E\left\{ {\sin {\varsigma _p}\sin {\varsigma _q}\cos ({\delta _p} + {\varsigma _p} - {\delta _q} - {\varsigma _q})} \right\} \\
\!\!\!\!= \frac{1}{4}({(E\{ \sin ({\delta _p} + 2{\varsigma _p})\} )^2} \!+\! {(E\{ \cos ({\delta _p} + 2{\varsigma _p})\} )^2} \\
\!\!\!\!+\! {(E\{ \sin{\!\delta _p}\}\! )^2} \!\!+\!\! {(E\{ \cos {\delta _p}\} \!)^2} \!\!-\!\! 2E\{ \sin ({\delta _p} \!\!+\! 2{\varsigma _p})\!\} E\{ \sin{\delta _p}\}  \\
- 2E\{ \cos ({\delta _p} + 2{\varsigma _p})\} E\{ \cos {\delta _p}\} ).
\end{array}
\!\!\!\!\!\!\!\!\!\!\label{Error-Omega-P-Expectation}
\end{equation}
Similar to the derivation process of $E\left\{ {\cos g_p} \right\}$ between Eq. (\ref{Error-Xi-P-E-COS}) and Eq. (\ref{Error-Xi-P-E-COS-ResultFinal}), we have
\begin{equation}
\!\!\begin{array}{l}
E\{ \cos ({\delta _p} + 2{\varsigma _p})\}  = \frac{{\sin (2\pi f{\tau _{\max }}) - \sin (2\pi f{\tau _{\min }})}}{{2\pi f({\tau _{\max }} - {\tau _{\min }})}}\\
{J_0}(\frac{{\gamma  - (2{C_1} + {F_1})}}{2}){J_0}(\frac{{\gamma  + 2{C_1} + {F_1}}}{2})\\
{J_0}(\sqrt {D_{n,h}^2 \!+\! D_{n,v}^2} \!-\! {D_{n,v}}){J_0}(\sqrt {D_{n,h}^2 \!+\! D_{n,v}^2} \!+\! {D_{n,v}}),
\end{array}
\!\!\!\label{Error-Omega-P-DeriveResult1}
\end{equation}
\begin{equation}
\!\!\begin{array}{l}
E\{ {\cos {\delta _p}}\}  = \frac{{\sin (2\pi f{\tau _{\max }}) - \sin (2\pi f{\tau _{\min }})}}{{2\pi f({\tau _{\max }} - {\tau _{\min }})}}{J_0}(\frac{{\upsilon  - {F_1}}}{2}){J_0}(\frac{{\upsilon  + {F_1}}}{2})\\
{J_0}(\sqrt {D_{n,h}^2 + D_{n,v}^2}  - {D_{n,v}}){J_0}(\sqrt {D_{n,h}^2 + D_{n,v}^2}  + {D_{n,v}}),
\end{array}
\!\!\!\label{Error-Omega-P-DeriveResult2}
\end{equation}
\begin{equation}
\!\!\begin{array}{l}
E\{ \sin ({\delta _p} + 2{\varsigma _p})\} = E\{ {\sin {\delta _p}}\} = 0.
\end{array}
\!\!\!\label{Error-Omega-P-SINResult}
\end{equation}
Therefore, the closed-form expression of $\mathop {\lim }\limits_{{P_1} \to \infty } \frac{\Lambda }{{{P_1} - 1}}$ is
\begin{equation}
\!\!\!\!\!\!\begin{array}{l}
\mathop {\lim }\limits_{{P_1} \to \infty } \frac{\Lambda }{{{P_1} - 1}} = \frac{(E\{ \cos ({\delta _p} + 2{\varsigma _p})\} - E\{ {\cos {\delta _p}}\})^2}{{4({K_R} + 1)}}\\
\ \ = \frac{{{{(\sin (2\pi f{\tau _{\max }}) - \sin (2\pi f{\tau _{\min }}))}^2}}}{{16{\pi ^2}{f^2}{{({\tau _{\max }} - {\tau _{\min }})}^2}({K_R} + 1)}}J_0^2(\!\sqrt {D_{n,h}^2 \!\!+\!\! D_{n,v}^2} \!\!-\!\! {D_{n,v}})\\
\ \ \ \ \ J_0^2(\!\sqrt {D_{n,h}^2 \!\!+\!\! D_{n,v}^2} \!\!+\!\! {D_{n,v}})\\
\ \ \ {({J_0}(\frac{{\gamma  - (2{C_1} + {F_1})}}{2}){J_0}(\frac{{\gamma  + 2{C_1} + {F_1}}}{2}) \!-\! {J_0}(\frac{{\upsilon  - {F_1}}}{2}){J_0}(\frac{{\upsilon  + {F_1}}}{2}))^2}.
\end{array}
\!\!\!\!\!\!\!\!\!\!\!\!\!\!\!\label{Error-Omega-P-Expectation-ResultFinal}
\end{equation}
Thus, Lemma \ref{Lemma3} is proved.

\subsection{Proof of Theorem \ref{Theorem2}}\label{Lemma5: error of no cross term}

In Theorem \ref{Theorem1}, we have proved that, for a LoS channel, the prediction MSE converges to zero, provided enough FA length and enough density of ports.
Evidently, the global optimal port of a LoS channel is a local optimal option for a multi-path channel.
Under the same conditions, for a multi-path channel, we may easily infer that the prediction MSE of the global optimal port is not larger than the prediction MSE of a local optimal port.
Therefore, $\left\| \varepsilon _n \right\|_2^2$ satisfies:
\begin{equation}
\begin{array}{l}
\!\!\!\!\mathop {\lim }\limits_{{N_t},\rho, P \to \infty } \left\| \varepsilon _n \right\|_2^2 \le 4\left\| {\sum\limits_{p = 1}^{{P_1}} {} {\alpha _p}{\beta _p}\sin {\varsigma _p}{e^{j(\frac{\pi }{2} + {\delta _p} + {\varsigma _p})}}} \right\|_2^2.
\end{array}
\!\!\!\!\label{Error-Asymptotically}
\end{equation}
By applying the inequality:
\begin{equation}
\begin{array}{l}
\left\| {{\cal{A}} + {\cal{B}}} \right\|_2^2 \le \left\| {\cal{A}} \right\|_2^2 + \left\| {\cal{B}} \right\|_2^2,
\end{array}
\!\!\!\!\label{Inequality-norm}
\end{equation}
we can derive
\begin{equation}
\!\!\!\!\!\!\!\begin{array}{l}
\!\!\!\!\!\!\!\!\mathop {\lim }\limits_{{N_t},\rho, P \to \infty }\!\! \left\| \varepsilon _n \right\|_2^2 \le {\cal U} = 4\sum\limits_{p = 1}^{{P_1}} {} \left\| {\alpha _p}{{\beta _p}\sin {\varsigma _p}{e^{j(\frac{\pi }{2} + {\delta _p} + {\varsigma _p})}}} \right\|_2^2\\
\!\!\!\!\!=\!\!\mathop {\lim }\limits_{ P_1 \to \infty }\!\!4\!\!\sum\limits_{p = 1}^{{P_1}} {}\!{\alpha _p^2} \beta _p^2{\sin ^2}{\varsigma _p} \!=\!\! \frac{2}{{{K_R} + 1}}(1 \!-\!\!\!\! \mathop {\lim }\limits_{ P_1 \to \infty }\!\!\frac{1}{{{P_1}}}\!\!\sum\limits_{p = 1}^{{P_1}} {} \cos(2{\varsigma _p}))\\
\!\!\!\!\!=\frac{2}{{{K_R} + 1}}(1 - E\{ \cos(2{\varsigma _p})\} ),
\end{array}
\!\!\!\!\!\!\!\!\!\!\!\!\!\!\!\label{Error-Asymptotically}
\end{equation}
where the closed-form expression of $E\{ \cos(2{\varsigma _p})\}$ is derived in Eq. (\ref{Sin2-closed-form}).
If $\mathop {\lim }\limits_{{N_t},\rho, P \to \infty } \left\| \varepsilon _n \right\|_2^2 = 4({\cal X} +{\cal Y}+{\cal Z} )$, we have
\begin{equation}
\!\!\!\!\!\!\!\!\begin{array}{l}
\mathop {\lim }\limits_{{P} \to \infty } \Lambda  = \mathop {\lim }\limits_{{P_1} \to \infty } \Lambda = \mathop {\lim }\limits_{{P_1} \to \infty } ( - 1 + \frac{1}{{\sqrt {{P_1}} }} - \frac{{\sqrt {{P_1}}  - 1}}{{\sqrt {{P_1}} ({P_1} - 2)}})\Omega \\ \ \ \ \ \ \ \ \ \ \ \approx \mathop {\lim }\limits_{{P_1} \to \infty } ( - \Omega  + \frac{\Omega }{{\sqrt {{P_1}} }}) = - \Omega  + {\cal X}.
\end{array}
\!\!\!\!\label{Error-equality}
\end{equation}
Due to $0 \le \mathop {\lim }\limits_{{N_t},\rho, P \to \infty } \left\| \varepsilon _n \right\|_2^2 \le \cal U$, we may obtain that if $\Lambda  <  - \Omega  + {\cal X}$, 
\begin{equation}
\begin{array}{l}
\!\!\!0 \le \!\!\!\mathop {\lim }\limits_{{N_t},\rho, P \to \infty } \!\!\!\left\| \varepsilon _n \right\|_2^2 \le \!{\min ( {4({\cal X} +{\cal Y}+{\cal Z} ),{\cal U} } )}.
\end{array}
\!\!\!\!\label{Error-inequality-less}
\end{equation}
As $\Lambda  > - \Omega  + {\cal X}$, 
\begin{equation}
\begin{array}{l}
\!\!\max (0,4({\cal X} +{\cal Y}+{\cal Z} )) \!\le \!\!\!\mathop {\lim }\limits_{{N_t},\rho, P \to \infty } \!\!\!\left\| \varepsilon _n \right\|_2^2 \!\le\! {\cal U}.
\end{array}
\!\!\!\!\label{Error-inequality-more}
\end{equation}
Thus, Theorem \ref{Theorem2} is proved.

\ifCLASSOPTIONcaptionsoff
  \newpage
\fi



%
\bibliographystyle{IEEEtran}
\bibliography{references.bbl}




\end{document}